\documentclass{article}
\usepackage{graphicx}
\usepackage{blindtext}
\usepackage[left=2.75cm, right=2.75cm]{geometry}
\usepackage{amsmath,amssymb,amsfonts}%
\usepackage{amsthm}%
\usepackage{mathrsfs}%
\usepackage{xcolor}%
\usepackage{textcomp}%
\usepackage{manyfoot}%
\usepackage{booktabs}%
\usepackage{algorithm}%
\usepackage{algorithmicx}%
\usepackage{algpseudocode}%
\usepackage{listings}%
\usepackage{euscript}
\usepackage{rotating}
\usepackage{xr}
\usepackage{capt-of}
\usepackage{booktabs}
\usepackage{tabularx}
\usepackage{caption}
\usepackage{subcaption}
\usepackage{hyperref}
\usepackage{cleveref}
\usepackage[mathlines]{lineno}
\usepackage{comment}

\renewcommand{\thetable}{S\arabic{table}}
\newcolumntype{C}{>{\centering\arraybackslash}X}
\newcommand{\AU}[3]{#1$^{#2}$ (\href{mailto:#3}{\texttt{#3}})}

\title{Computational Mapping of Reactive Stroma in Prostate Cancer Yields Interpretable, Prognostic Biomarkers}
\usepackage[sorting=none,maxbibnames=10,backend=biber]{biblatex}
\begin{document}
\author{%
\begin{minipage}{0.95\textwidth}\centering\sloppy
\AU{Mara Pleasure}{1,2,3}{mpleasure@ucla.edu},\;
\AU{Ekaterina Redekop}{1,2,4}{eredekop@ucla.edu},\;
\AU{Dhakshina Ilango}{1}{dhakshina@ucla.edu},\;
\AU{Zichen Wang}{1,2,4}{zcwang0702@ucla.edu},\;
\AU{Vedrana Ivezic}{1,2,3}{vivezic@ucla.edu},\;
\AU{Kimberly Flores}{5}{kimberlyflores@mednet.ucla.edu},\;
\AU{Israa Laklouk}{5}{ILaklouk@mednet.ucla.edu},\;
\AU{Jitin Makker}{5}{JMakker@mednet.ucla.edu},\;
\AU{Gregory Fishbein}{5}{GFishbein@mednet.ucla.edu},\;
\AU{Anthony Sisk}{5}{asisk@mednet.ucla.edu},\;
\AU{William Speier}{1,2,3,4}{speier@ucla.edu},\;
\AU{Corey W. Arnold}{1,2,3,4,5*}{cwarnold@ucla.edu}\\[0.75em]
\small
$^{1}$ Biomedical AI Research Lab, University of California, Los Angeles \quad
$^{2}$ Department of Radiology, University of California, Los Angeles \quad
$^{3}$ Bioinformatics Interdepartmental Program, University of California, Los Angeles \\
$^{4}$ Department of Bioengineering, University of California, Los Angeles \quad
$^{5}$ Department of Pathology \& Laboratory Medicine, University of California, Los Angeles\\[0.35em]
$^{*}$\textit{Corresponding author:} Corey W. Arnold, \href{mailto:cwarnold@ucla.edu}{cwarnold@ucla.edu}
\end{minipage}
}
\date{}
\pagestyle{empty}

\maketitle

\section*{Abstract}
Current histopathological grading of prostate cancer relies primarily on glandular architecture, largely overlooking the tumor microenvironment. Here, we present PROTAS, a deep learning framework that quantifies reactive stroma (RS) in routine hematoxylin and eosin (H\&E) slides and links stromal morphology to underlying biology. PROTAS-defined RS is characterized by nuclear enlargement, collagen disorganization, and transcriptomic enrichment of contractile pathways. PROTAS detects RS robustly in the external Prostate, Lung, Colorectal, and Ovarian (PLCO) dataset and, using domain-adversarial training, generalizes to diagnostic biopsies. In head-to-head comparisons, PROTAS outperforms pathologists for RS detection, and spatial RS features predict biochemical recurrence independently of established prognostic variables (c-index 0.80). By capturing subtle stromal phenotypes associated with tumor progression, PROTAS provides an interpretable, scalable biomarker to refine risk stratification.
\bigskip
\subsubsection*{Keywords}
\textit{Computational Pathology, Reactive Stroma, Prostate Cancer}

\section{Introduction}
In 2024, prostate cancer was the most commonly diagnosed cancer among U.S. men and the second leading cause of cancer death \cite{siegel2024cancer}. Clinical management of prostate cancer begins with histological grading of biopsy specimens using the Gleason system, created in 1966 and refined in 2014 by the International Society of Urological Pathology (ISUP) \cite{delahunt2012gleason, srigley2016one, epstein20162014}. This system relies on glandular morphology to stratify patients into risk categories that guide treatment decisions, from active surveillance for low-risk disease to definitive intervention for higher cancer grades \cite{hirz2023dissecting, humphrey2017histopathology}.

Despite improvements in risk stratification and prognostic accuracy, substantial heterogeneity persists. Prostate cancer is heterogeneous, with patients often exhibiting different responses to treatment and patterns of progression even when sharing the same Gleason pattern or ISUP grade group \cite{srigley2016one, eastham2022clinically}. This variability reflects two key limitations of current grading paradigms, (1) high inter-observer variability and (2) an exclusive focus on glandular architecture that overlooks the tumor microenvironment (TME) \cite{bulten2020automated, humphrey2017histopathology}. Since the introduction of the Gleason grading system, numerous studies have highlighted the critical role of the TME and surrounding host tissue in patient outcomes. Tumor progression across various solid malignancies is now understood to be driven by complex interactions between cancer cells and their microenvironment, which includes stromal cells, immune cells, and extracellular matrix (ECM) components, all of which coexist and contribute to different patient outcomes \cite{goncalves2015key}.

In invasive cancer, the surrounding stroma may undergo alterations characterized by ECM remodeling, cancer-associated fibroblast (CAF) infiltration, increased inflammation, and growth factor secretion \cite{goncalves2015key, nederlof2022spatial}. Collagen remodeling and degradation facilitate angiogenesis and tissue invasion, processes central to tumor progression \cite{goncalves2015key}. Evidence from breast, colorectal, and pancreatic ductal cancers demonstrates that stromal composition and organization are important prognostic indicators \cite{bulle2020beyond, lu2023tumor}. In breast cancer, the ratio of tumor-to-stroma has prognostic value, while in pancreatic ductal cancer, high collagen fiber alignment is predictive of poor outcomes \cite{bhattacharjee2022invasive, drifkahighly}.

In prostate cancer, reactive stroma (RS), characterized by increased ECM deposition, CAF presence, and inflammatory cell infiltration, has emerged as a potential prognostic marker \cite{ayala2003reactive, de2017histologic, silva2015characterization}. Ayala et al. first demonstrated that RS identified by Masson's Trichrome staining was an independent predictor of reduced recurrence-free survival \cite{ayala2003reactive}. Subsequent work from the same group, De Vivar et al. introduced a three-tiered grading system for identifying RS on hematoxylin \& eosin (H\&E) stained slides, with the highest grade, RSG3, found to be predictive of biochemical recurrence independent of Gleason grade \cite{de2017histologic}. Transcriptomic analyses have linked these stromal changes to wound-healing phenotypes and growth signaling pathways \cite{dakhova2014genes}, while CAF proportions predict poor outcomes, metastasis, and therapeutic resistance \cite{bonollo2020role, blom2019fibroblast}.

Despite this evidence, clinical adoption of stroma assessment remains limited \cite{keikhosravi2020non, ruder2022development, yanagisawa2007stromogenic, ding2024quantification}. Recognition of the stroma’s role in prostate cancer progression has been gradual, often restricted to specialized sites with internal grading systems \cite{keikhosravi2020non, ruder2022development, yanagisawa2007stromogenic, ding2024quantification}. Most RS identification methods rely on expert pathologists with specialized training, immunohistochemistry, or manual analysis, approaches that are labor intensive, subjective, and difficult to scale \cite{ayala2003reactive, ruder2022development, yanagisawa2007stromogenic, de2017histologic, ayala2011determining, tuxhorn2001reactive, ding2024quantification}. Even quantitative H\&E-based efforts have relied on expert-labeled internal datasets and proprietary software, limiting reproducibility and broader implementation \cite{ruder2022development}. Critically, we lack open, quantitative tools that learn from routine H\&E without specialized labeling that systematically characterize how RS varies by histological context and spatial proximity to tumor \cite{ayala2003reactive, ruder2022development, yanagisawa2007stromogenic, de2017histologic, ayala2011determining}. Objective, quantitative approaches are needed to identify and characterize stromal changes on routine H\&E and relate them to diagnosis and outcomes, and enable whole-prostate spatially aware biomarker discovery.

To address this translational gap, we reasoned that deep learning could standardize this subjective phenotype, transforming RS from a qualitative observation into a quantitative, reproducible biomarker. Unlike prior attempts that relied on small, private datasets with expert labels or specialized stains \cite{ruder2022development}, we sought to develop a robust, H\&E framework capable of characterizing stromal landscapes at scale and links them directly to clinical outcomes.

Here, we introduce PROTAS, a deep learning framework trained on 255 whole-mount radical prostatectomy (RP) specimens from patients treated at the University of California, Los Angeles (UCLA), and validated on the independent multi-institutional PLCO cohort \cite{zhu2013prostate}. Using PROTAS, we map spatial patterns of RS across grade groups and show that slide-level RS features improve prediction of biochemical recurrence (BCR) beyond clinical baselines such as the UCSF-CAPRA score \cite{cooperberg2005university}. We further anchor PROTAS' predictions in biology by linking PROTAS-defined RS to interpretable morphological changes and transcriptomic signatures, confirming the model specifically identifies a contractile and ECM-remodeling phenotype. We benchmarked PROTAS against four board-certified pathologists for RS identification and extend the framework to prostate biopsy specimens, demonstrating feasibility for RS detection in smaller diagnostic specimens with implications for reducing underdiagnosis.

\section{Results}

\subsection{PROTAS identifies reactive stroma in radical prostatectomy tissues
}

To characterize the spatial distribution of Reactive Stroma (RS), we required a scalable method to identify RS on routine H\&E slides. Since no public datasets delineate RS, we developed PROTAS using a spatially defined hypothesis: stroma adjacent to cancer (RS-positive) is more likely to exhibit distinct morphology than stroma distant from cancer (RS-negative), even when these differences are not readily apparent by human inspection (\nameref{sec:methods}, Figure.\ref{fig:1}(b)).

We trained PROTAS on 255 whole-mount whole-slide image (WSI) RP specimens from UCLA, with external validation on the PLCO cohort. Whole-mount specimens enabled identification of dominant tumor foci and accurate assignment of RS-positive labels, while also providing sufficient surrounding normal tissue to delineate RS-negative regions far from any cancer focus. To ensure the model learned stromal features rather than epithelial patterns, we utilized semantic segmentation to exclude glandular and neoplastic epithelium, retaining stromal-only patches for model training and downstream analyses.

On the held-out test set (n = 64), PROTAS effectively discriminated between tumor-adjacent and distant stroma (AUROC = 0.85, Sensitivity = 0.83, Specificity = 0.75), consistent with a morphologically distinct stromal phenotype recognizable by deep learning (Figure \ref{fig:1}c). On the independent PLCO cohort (n = 71), the model maintained robust discrimination (AUROC = 0.80) despite differences in tissue processing and scanning, demonstrating that the PROTAS-defined RS signature is generalizable across diverse clinical settings.

\subsection{Reactive stroma patterns vary by grade group and MRI risk category}

We next applied PROTAS to whole-mount RP specimens to quantify the spatial extent of RS and investigate its relationship with established risk factors.

\subsubsection{RS features differ between low- and high-grade tumors}
To compare RS patterns quantitatively, we stratified patients into two groups, $\leq$GG2 (Grade Groups 1-2, n = 98) or $\geq$GG3 (Grade Groups 3-5, n = 151). For each patient, we computed various summary slide-level features capturing RS patterns from regions outside any histologically confirmed cancer (\nameref{sec:methods}). Between-group differences were assessed using Wilcoxon rank-sum test; p-values were Bonferroni adjusted for multiple comparisons (family-wise $\alpha = 0.05$). 

Patients with $\geq$GG3 tumors showed significantly greater RS burden, with a higher fraction of high-probability RS patches (adjusted $p < 0.001$), higher entropy within RS-positive patches (adjusted $p < 0.001$), larger RS-hotspot graph diameters (adjusted $p < 0.001$), and a higher maximum RS rate across 1-mm distance bins (adjusted $p< 0.001$). Conversely, $\leq$GG2 cases showed fewer hotspot regions (adjusted $p < 0.001$), but higher hotspot Laplacian mean (adjusted $p < 0.001$), consistent with more spatially isolated RS areas (Figure \ref{fig:2}(a,c), Supplementary Table \ref{tab:s1}).

\subsubsection{RS features associate with preoperative MRI risk stratification}
We next examined whether RS patterns correspond to preoperative multiparametric MRI (mpMRI) findings using the Prostate Imaging Reporting and Data System (PI-RADS) categories, a five-point scale for lesion suspicion. \cite{gupta2020pi} We stratified patients by their PI-RADS scores into two categories: 2-4 (low/intermediate risk, n = 75) versus 5 (high risk, n = 93), and applied the same slide-level feature analysis. 

PI-RADS 5 cases exhibited lower hotspot Laplacian mean (adjusted $p < 0.05$), a greater number of hotspot-to-hotspot connections (adjusted $p < 0.01$), and a higher maximum RS rate across 1-mm distance bins (adjusted $p < 0.05$). However, fewer features reached significance between PI-RADS groups compared to grade group stratification, suggesting RS patterns are strongly associated with histological grade rather than imaging-based risk scores (Figure \ref{fig:2}(b,c), Supplementary Table \ref{tab:s2}).

\subsection{RS features improve biochemical recurrence prediction beyond clinical baselines}

To evaluate the prognostic utility of RS, we assessed whether slide-level RS features from regions outside tumor boundaries could predict biochemical recurrence (BCR) using Cox proportional hazards models (\nameref{sec:methods}). We compared models based on RS feature to the UCSF-CAPRA score, and to tumor characteristics (e.g., tumor size, multifocality, epithelial-stroma ratio). Following Cooperberg et al., we excluded patients with pathologic TNM stage (AJCC 7th/8th edition) $\geq$ T3b, $<$ 2 post-RP serum prostate-specific antigen (PSA) measurements, secondary treatment within 6 months of RP, or missing clinical variables \cite{cooperberg2005university}. BCR was defined as two consecutive serum PSA measurements $\geq$ 0.2ng/ml after RP or patients who received secondary prostate cancer treatment $>$ 6 months post-RP. The final cohort consisted of 164 patients, with a median long-term follow-up of 76.3 months and a median time to BCR of 23.7 months.

We fit separate Cox models using RS features alone, UCSF-CAPRA alone, and tumor features alone, and then evaluated pairwise combinations of feature sets. Discrimination was assessed using the concordance index (c-index) and time-dependent AUC at 12, 24, and 60 months (\nameref{sec:methods}). 

Models based on RS features alone (c-index = 0.711, 95\% CI 0.64-0.78) and tumor features alone (c-index = 0.736, 95\% CI 0.67-0.8) both outperformed the UCSF-CAPRA alone model (c-index = 0.663, 95\% CI 0.59-0.73). Notably, combining RS with tumor features achieved the strongest performance (c-index = 0.804, 95\% CI 0.75-0.85; AUC@12 months = 0.84, 95\% CI 0.76-0.91; AUC@24 months = 0.877, 95\% CI 0.81-0.93; AUC@60 months = 0.835, 95\% CI 0.75-0.91), demonstrating that stromal and tumor features provide complementary prognostic information. Adding CAPRA to the RS + tumor features yielded a modest gain at 12 months but otherwise performed similarly or slightly lower (c-index = 0.795, 95\% CI 0.73-0.85; AUC@12 months = 0.864, 95\% CI 0.77-0.94; AUC@24 months = 0.858, 95\% CI 0.79-0.92; AUC@60 months = 0.827, 95\% CI 0.74-0.9), (Figure \ref{fig:3}(a), Supplementary Table \ref{tab:s3}, Supplementary Table \ref{tab:s4}). 

To identify which RS features most strongly associate with outcomes, we examined individual features in multivariable Cox models with clinical adjustment (e.g., PSA, age, TNM stage, GG) (Supplementary Table \ref{tab:s5}). Entropy of RS probabilities across 1-mm distance bins showed the most consistent association with poorer prognosis (HR 3.3-6.7), remaining significant across all adjustment sets except those controlling for tumor features alone. RS edge-to-core ratio, a feature that captures RS distribution from the prostate periphery to the tumor border, also showed risk-increasing associations in several adjustment sets (HR 3.0-5.8), whereas other features (e.g., hotspot Laplacian roughness) showed moderate but less consistent effects. These findings indicate that spatial patterns of RS, in particular distance-dependent and hotspot distribution, provide independent prognostic information beyond standard clinical variables. 

Stratifying patients into quartiles based on RS-derived prognostic scores yielded clear separation of Kaplan-Meier curves across all models, with the RS + tumor feature combination yielding the strongest separation (Figure \ref{fig:3}(c), Supplementary Table \ref{tab:s6}).

\subsection{PROTAS learns a complex, multi-scale phenotype beyond simple morphology}
To interpret the tissue characteristics underlying PROTAS predictions, we quantified nuclear and collagen features that differentiate RS-positive from RS-negative tissue, providing interpretable correlates of the deep learning model's decisions (\nameref{sec:methods}). 

\subsubsection{Nuclear and Collagen Signatures of Reactivity}
After controlling for inter-patient variability (patient fixed effects), we found RS-positive regions exhibited systematic differences in stromal morphology and extracellular matrix (ECM) organization. RS-positive tissue showed nuclear atypia, characterized by larger, more solid nuclei (OR = 1.43, 95\% CI 1.35-1.51, adj. p $<$ 0.0001) with higher intensity variability (Figure \ref{fig:4}b). Consistent with prior work linking nuclear orientation to cancer severity and clinical outcomes, RS-positive tissue also showed increased variability in nuclear orientation (OR = 1.21, 95\% CI 1.15-1.27, adjusted $p < 0.0001$) (Figure \ref{fig:4}(b))\cite{lu2018nuclear}. 

We next probed the ECM using a generative model to synthesize collagen fiber maps. RS-positive tissue demonstrated a disorganized remodeling phenotype: increased overall collagen volume (OR = 1.27, 95\% CI 1.20-1.36, adj. p $<$ 0.0001), but decreased fiber alignment (OR = 0.81, 95\% CI 0.76-0.85, adj. p $<$ 0.0001) and shorter fiber lengths (OR = 0.82, 95\% CI 0.77-0.87, adj. p $<$ 0.0001). Together these findings indicate that PROTAS identifies stromal regions consistent with RS-associated nuclear and matrix remodeling.

\subsubsection{Deep learning features outperform handcrafted morphological features for RS classification}
While individual morphological features were significantly associated with RS, they were insufficient for accurate classification on their own. We trained multiple classifiers (Random Forest, XGBoost, MLP) using the engineered nuclear and collagen features. The best performing handcrafted model achieved an AUROC of 0.728 (vs. PROTAS 0.85). This performance gap suggests that deep learning captures stromal patterns beyond those represented by predefined morphological features (Figure \ref{fig:4}a).

\subsubsection{Feature Attribution}
SHapley Additive exPlanations (SHAP) analysis revealed that the handcrafted-feature models integrate signals from both cellular and matrix compartments. The top 15 features showed consistent directional effects across RS-positive and RS-negative patches, with individual contributions of each feature visualized for 100 randomly sampled RS-positive and RS-negative patches (Figure \ref{fig:4}(c,d)).

\subsection{PROTAS provides objective reactive stroma identification with greater consistency than expert pathologists}
To benchmark PROTAS against expert visual assessment, we compared the model's predictions to four board-certified pathologists (three specialized in genitourinary pathology). The pathologists independently reviewed 100 randomly selected stroma-only patches (50 RS-positive, 50 RS-negative) from the test set and categorized each as “abnormal” or “normal” stroma, recorded their confidence level, and described any visual cues they used to make their decision (cytoplasm color, nuclear shape and color, extracellular matrix, other) (\nameref{sec:methods}; Supplementary Figure \ref{supp_fig3:1}). 

On this random subset, PROTAS substantially outperformed all individual raters across all three metrics (accuracy = 0.7, recall = 0.88, precision = 0.64). The best-performing pathologist achieved an accuracy = 0.59, recall = 0.49, and precision = 0.60. Majority-vote ensembles showed similar or lower performance: all four pathologists (accuracy = 0.55, recall = 0.51, precision = 0.54), and prostate-specialized pathologists only (n = 3, accuracy = 0.62, recall = 0.49, precision = 0.65) (Supplementary Table \ref{tab:s10}). 

Agreement analysis further highlighted the difficulty of visually identifying subtle stromal phenotypes on routine H\&E. Cohen’s Kappa analysis revealed minimal agreement between PROTAS and pathologists (Model vs. P1 $\kappa$ = 0.143, Model vs. P2 $\kappa$ = 0.045, Model vs. P3 $\kappa$ = -0.023, Model vs. P4 $\kappa$ = 0.042), and poor inter-rater agreement among pathologists themselves ($\kappa$ = 0.013-0.279)(Supplementary Table \ref{tab:s11}). These findings underscore the challenge of manually identifying subtle stromal changes on routine H\&E and demonstrate PROTAS' ability to provide objective, reproducible RS assessment that is not readily achievable through visual evaluation.

\subsection{Gene expression reveals differences between RS-positive and RS-negative patches}

To test whether PROTAS-derived RS is associated with distinct transcriptional programs, we aligned model predictions with bulk RNA-sequencing data from the TCGA-PRAD cohort (n = 317 patients). We extracted stroma-only patches, computed PROTAS predictions, and generated slide-level RS features using our established pipeline (\nameref{sec:methods}). Patients stratified as RS-rich ($\hat{p}\geq 0.75$) by PROTAS exhibited a distinct transcriptomic profile characterized by the upregulation of contractile and smooth muscle pathways, including myofibril assembly, sarcomere organization, and striated muscle development (FDR $<$ 0.05, all adjusted $p < 0.01$) (Figure \ref{fig:5.2.1}) \cite{kuleshov2016enrichr}.  These findings align with cancer-associated fibroblast biology and contractile phenotypes of ECM remodeling.

In contrast, differential expression between grade groups ($\leq$GG2 vs. $\geq$GG3) was dominated by cell-cycle and mitotic processes (e.g., mitotic nuclear division (Figure \ref{fig:5.1}, Figure \ref{fig:5.2.2}). This distinct enrichment profile support PROTAS-derived RS as a transcriptional axis that is separable from proliferative programs associated with tumor grade.

We next evaluated these associations at higher spatial resolution using spatial transcriptomics (Visium) from the HEST dataset \cite{jaume2024hest}. By applying PROTAS to classify specific tissue spots (55$\mu m$ diameter) as RS-positive or RS-negative, we identified 1,973 differentially expressed genes (FDR $<$ 0.05). Several genes with established roles in prostate cancer progression were significantly enriched in RS-positive regions. Notably, \textit{TGFB1$|$1}(Hic-5), a paxillin family member implicated in metastasis and reactive stroma \cite{heitzer2007hic}, was consistently differentially expressed (Fisher-adjusted $p$ $<$ 0.0001), as was the protease \textit{KLK4} (Fisher-adjusted $p$ $<$ 0.001). Additional prostate cancer-associated markers, including \textit{CD24}, \textit{TSPAN1}, \textit{SPON2}, and \textit{FN1}, were also differentially expressed in PROTAS-positive regions. These multi-scale genomic associations confirm that PROTAS acts as a reliable morphological surrogate for a specific, biologically aggressive stromal phenotype.

\subsection{PROTAS detects reactive stroma in prostate biopsy cores through domain adaptation
}
Prostate biopsy cores are routinely used for diagnosis and treatment planning; however, they provide limited tissue sampling, which can lead to diagnostic uncertainty. To evaluate whether PROTAS could detect RS on biopsy cores, we analyzed 3,706 biopsy cores from 179 patients from our institution using a stromal-filtering and labeling approach adapted from our RP pipeline (\nameref{sec:methods}).

We curated an initial biopsy dataset for zero-shot testing evaluation by labeling negative patches as all stromal regions from benign-only biopsies, and positive patches as stromal regions from high-grade (Gleason pattern 4-5) cores. When applied to this test set, the RP-trained model performed poorly on biopsies (AUROC = 0.49, accuracy = 0.48, sensitivity = 0.41, specificity = 0.58), consistent with a substantial domain shift. Unsupervised UMAP visualization of patch embeddings confirmed clear separation between RP and biopsy specimens (Supplementary Figure \ref{supp_fig4:1}).

To mitigate domain shift, we incorporated biopsy data into our training procedure and used domain-adversarial adaptation to enforce latent-space alignment between RP and biopsy patches (\nameref{sec:methods}; Figure \ref{fig:6}) \cite{ganin2016domain}. Given the limited tissue in biopsies and difficulty defining spatial proximity to tumor, we refined our prior biopsy labeling strategy to incorporate information based on tumor content thresholds: cores with $\geq$50\% vs. $\geq$70\% for RS-positive labels. The model trained with $\geq$70\% tumor cores achieved superior performance (AUROC = 0.821, sensitivity = 0.702, specificity = 0.803, AUPRC = 0.794) compared to the model trained on cores with $\geq$50\% tumor (AUROC = 0.783, sensitivity = 0.73, specificity = 0.73, AUPRC = 0.788). This substantial improvement over zero-shot application ($\Delta$AUROC = +0.33) demonstrates that domain adaptation enables robust RS detection in biopsies.

\section{Discussion}
Prostate cancer risk stratification is still largely anchored to the Gleason grading system, a framework focused primarily on epithelial glandular architecture. While powerful, this tumor-centric paradigm overlooks the tumor microenvironment. Here, we show that quantitative characterization of reactive stroma (RS) from routine H\&E provides prognostic information that is complementary to, and in several analyses independent of, standard clinicopathologic variables.

A critical barrier to the clinical adoption of stromal biomarkers has been the lack of reproducible standards. As our blinded reader study reveals, even board-certified pathologists struggle to agree on the presence of RS (inter-rater $\kappa$ $<$ 0.3), highlighting the difficulty of operationalizing RS detection by visual inspection alone. PROTAS addresses this bottleneck by converting a subjective qualitative observation into an standardized quantitative metric. PROTAS offers a standardized framework for H\&E slides that allows for stromal profiling without the need for specialized stains or expensive molecular assays.

Multiple orthogonal analyses support that PROTAS is capturing a meaningful biological pattern rather than a non-specific visual artifact. Across bulk transcriptomic analyses, PROTAS-defined RS was associated with enrichment activation and extracellular matrix (ECM) remodeling. These signatures were separable from grade-associated proliferation and cell-cycle programs, suggesting that PROTAS-derived RS features reflect a distinct axis of tumor biology. Spatial transcriptomic analyses further localized this program to PROTAS-positive regions and identified enrichment of markers implicated in stromal remodeling and prostate cancer progression including \textit{TGFB1$|$1} (Hic-5) and \textit{KLK4}. This is further supported by our interpretable morphological analysis, which identified a disorganized remodeling signature (high collagen volume, low alignment) consistent with extracellular remodeling and prior findings of RS \cite{cooperberg2005university, nederlof2022spatial}.

PROTAS also revealed that RS extent and spatial organization vary with histologic disease severity. Higher-grade tumors exhibited more RS hotspots, greater spatial heterogeneity (measured by distance-bin entropy), and elevated maximum RS probabilities near tumor boundaries. RS signal also correlated to mpMRI imaging, with several features demonstrating significant differences between highly suspicious lesions (PI-RADS 5) and lower-risk PI-RADS categories. Several of these features captured RS texture and spatial features, suggesting that the correlation was not based solely on tumor size.

The clinical relevance of these stromal signals is underscored by their prognostic value. In a cohort with long-term follow-up, PROTAS-derived features improved biochemical recurrence prediction compared with clinical baselines such as the UCSF-CAPRA score. Notably, spatial descriptors of RS organization, specifically the entropy and edge-to-core gradients drove this predictive performance. This suggests that patients with identical Gleason scores may harbor vastly different risks based on the spatial organization of their stroma. Finally, we demonstrated the translational feasibility in the diagnostic setting. By using domain-adversarial training to bridge the domain gap between resections and needle biopsies, we showed that RS stratification is possible at the point of initial diagnosis. 

There are several limitations of this study. First, based on prior work and tumor microenvironment dynamics, our study design assumed that RS exists proximal to high-grade cancer. Our approach provided a conservative baseline, but alternate definitions are possible and may capture complementary biology. Second, prognostic modeling was retrospective and single-center; prospective external validation is essential to assess the clinical utility of RS features across scanners, clinical centers, and patient populations. Third, while our synthetic and spatial transcriptomic analyses provide strong mechanistic evidence, they are limited by sample size and the generative nature of the imaging data. 

PROTAS offers an objective, reproducible method to quantify stromal biology from existing H\&E slides, complementing conventional grading systems. Integration into digital pathology workflows could provide pathologists with stromal risk estimates, alongside standard assessments, potentially refining risk stratification within grade groups. Overall, our results indicate that quantifying RS from routine pathology is feasible and yields interpretable, prognostically relevant biomarkers that are distinct from tumor-centric features. 

Beyond clinical use, PROTAS provides a research framework for investigating stromal remodeling. Researchers can apply PROTAS to existing prostate tissue specimens to derive RS features and incorporate them into prognostic models or multi-omic association studies. Future work should evaluate PROTAS in larger, prospective cohorts and assess its integration with genomic and transcriptomic biomarkers. Further exploration in biopsy specimens represents an important direction, as early detection of stromal remodeling in diagnostic biopsies could inform prognosis and treatment planning. Ultimately, quantifying RS from routine pathology expands our understanding of prostate cancer biology and offers new opportunities for personalized risk assessment.

\section{Methods}\label{sec:methods}
\subsection{Dataset Preparation}
\subsubsection{UCLA Radical Prostatectomy (RP) Dataset}
To establish high-quality spatial labels for RS, we implemented stringent quality control on whole-mount radical prostatectomy (RP) specimens to avoid false positives. Starting with our internal UCLA database of WSIs and pathology reports, we matched RP pathology reports to whole-mount WSIs and retained only cases with slide-specific pathology details in the report. A tumor segmentation model previously developed in our laboratory was applied to the WSIs to generate tumor masks, and cases showing discrepancies between the report and mask were excluded. For patients with multiple whole-mount WSIs that matched, we chose the slide with the largest dominant foci tumor burden. This yielded 255 patients, each with one corresponding whole-mount RP whole-slide image (WSI) per patient stained with hematoxylin and eosin (H\&E). 

To isolate stroma-only regions, we applied a two-stage segmentation pipeline. First, a gland-segmentation model classified glandular structures \cite{wang2023deep}; second, CellViT \cite{horst2024cellvit} segmented and classified individual nuclei. We extracted 256$\times$256 pixel patches (40$\times$) and excluded any patch containing epithelial or neoplastic nuclei. The remaining stroma-only patches were stain-normalized and embedded using the UNI foundation model \cite{chen2024towards} into a 1024-dimensional feature vector.

For RS-labeling, we defined RS-positive as stroma residing within the tumor bed of clinically significant cancer (GG $\geq$ 2). RS-negative was defined as stroma located $\geq$ 10mm from any tumor focus (including non-dominant foci) in cases with low-grade dominant tumors ($\leq$ GG2). This conservative spatial filtering minimized label noise by avoiding the transition zone between tumor and normal tissue. The dataset was split into patient-level training (n = 169), validation (n = 29), and test (n = 64) sets.

\subsubsection{External validation: PLCO}
For external validation, we used the PLCO dataset, which provides some quarter- and half-prostate specimens rather than whole-mount slides. Since there are no other publicly available whole-mount WSI datasets, we wanted to use the large tissues available in PLCO to our advantage. We started by selecting samples from patients with only very low-grade (GG1) or very high-grade (GG4/GG5) cancers. We then further limited this set of samples to only the cases that contained tumor-annotation marker masks specifying tumor location and sufficient non-tumor tissue ($\geq$ 10mm of distance from the tumor boundaries). After our filtering, we ended up with a dataset of 71 samples, with a grade-group breakdown of 41 GG1, 13 GG4, and 17 GG5. Stroma extraction and feature embedding follow the same pipeline as the internal RP dataset.

\subsubsection{UCLA Biopsy Dataset}
Needle biopsies sample limited tissue, precluding the spatial proximity approach used for whole-mount RP specimens. We therefore defined the RS classes based on tumor content. For negative examples, we identified patients with entirely benign biopsy findings (3,029 slides, n = 75 patients). For positive examples, we selected cores with GG $\geq$ 3 and $\geq$50\% tumor involvement. All slides were tiled at 40$\times$ into 256$\times$256-pixel patches and processed with CellViT to segment nuclei and classify them as epithelial versus endothelial/inflammatory. Patches containing any epithelial (glandular) nuclei were excluded; the remaining stroma-only patches were retained and labeled negative. Patches were extracted, normalized, and embedded using the same pipeline as RP specimens. For the stricter labeling strategy reported in Results, we increased the tumor content threshold to $\geq$70\% for positive class definition.

\subsubsection{Ethics Approval}
All human data in this study were collected and analyzed under protocols approved by the Institutional Review Board of the University of California, Los Angeles (IRB-16-1361). The requirement for informed consent was waived due to the retrospective nature of the study. External data from the Prostate, Lung, Colorectal, and Ovarian (PLCO) Cancer Screening Trial were obtained under the PLCO data use agreement.

\subsection{Model Training \& Evaluation}
PROTAS consists of a frozen UNI feature encoder followed by a two-layer trainable Multi-Layer Perceptron (MLP). To mitigate the noise inherent in spatially-derived labels, we introduced two regularization techniques into training:
\begin{itemize}
    \item Severity-aware label smoothing: We assigned soft targets based on the Gleason score (e.g. target = 0.90 for Gleason 4+5; 0.75 for 4+3).
    \item Confidence Penalty: We added an entropy regularization term ($\alpha  = 0.5$) to the loss function to penalize overconfident predictions \cite{pereyra2017regularizing}.
\end{itemize}
The model was trained for 100 epochs using the Adam optimizer. At inference we quantified predictive uncertainty using Monte Carlo dropout (100 stochastic forward passes) \cite{gal2016dropout}.

\subsection{Heatmap Visualization}
To visualize stromal patterns, we generated continuous probability heatmaps by spatially aggregating patch-level predictions. Patches were downsampled to a target resolution and overlapping contributions were smoothed using a running average to create a seamless probability surface (detailed algorithm in Supplementary Methods). 

\subsection{Slide-Level Quantitative Features}
To transform these visual patterns into quantitative biomarkers we extracted a comprehensive set of stromal landscape features from regions outside the tumor boundary. This ensured we captured the field effect of the tumor rather than the tumor itself.
\begin{enumerate}
\item Basic Statistics: We computed the mean, variance, and entropy of RS probabilities across the slide to quantify overall stromal activation.
\item Spatial Hotspots: We identified contiguous regions of high-confidence RS ($\geq$ 75) and characterized their geometry. We calculated metrics such as hotspot density (regions per area), spatial dispersion (variance of pairwise distances), and roughness (Laplacian of the probability surface) to quantify how "patchy" or "smooth" the stromal reaction appeared.
\item Graph Topology: To measure connectivity, we constructed a spatial graph connecting RS hotspots with a defined radius ($\delta = 2000$ pixels). We compute graph-theoretic metrics including graph diameter and clustering coefficient to determine if RS presented as isolated pockets or a unified, connected network.
\item Distance-from-Cancer Gradients: To quantify the extent of reactivity, we binned stroma patches by their distance from the tumor boundary (1mm increments). We calculated the decay rate of RS probability as a function of distance, capturing how far the reactive stroma extends into normal tissue.
\item Conventional Morphological Controls: For comparison, we extracted standard tumor shape features (e.g. tumor area, solidity, perimeter) and tissue composition ratios (cancer-to-stroma ratio).
\end{enumerate}

(Full mathematical definitions for spatial and graph features are provided in Supplementary Methods).

\subsection{Cohort and Outcome Definition}
We performed a comprehensive chart review of UCLA patients undergoing Radical Prostatectomy (RP) to curate a prognostic cohort. Biochemical Recurrence (BCR) was defined as two consecutive post-RP serum PSA measurements $\geq$ 0.2 ng/ml or the initiation of secondary treatment $>$ 6 months post-RP.
\subsection{Survival Analysis}
We constructed Cox Proportional Hazards (CPH) models to predict time-to-BCR. Models were fit using the lifelines package. Discrimination was assessed using the concordance index (c-index) and time-dependent cumulative/dynamic AUC at 12, 24, and 60 months. To ensure robustness, all performance metrics and 95\% confidence intervals were estimated via bootstrap resampling (5,000 replicates).

\subsection{Feature Selection and Transformation}
To prevent overfitting, we applied a strict preprocessing pipeline. Features were first filtered for collinearity (Pearson's $r \geq 0.9$). Remaining features underwent distributional transformation (Yeo-Johnson or log-transform) and standardization. We fit a base CPH model and dropped features with low coefficient magnitudes and large standard errors. Features with high-variance were discretized into quantile-based bins. (See Supplementary Methods for detailed feature engineering steps).

\subsection{Confounder Analysis}
To assess robustness to clinical confounding, we re-fit CPH models including the RS feature set jointly with one clinical variable at a time and recorded changes in hazard ratios and p-values. (See Supplementary Methods for list of confounders).
 
\subsection{Kaplan-Meier Analysis}
For Kaplan-Meier analysis, we computed each model’s linear predictor (partial hazard) and stratified patients into four quartiles defined over the full cohort. Separate Kaplan-Meier curves were fit for each stratum (via lifelines), and survival distributions were compared using the log-rank test. Resulting p-values were adjusted for multiple comparisons with the Benjamini-Hochberg procedure.

\subsection{Interpretable Feature Analysis and Comparison}

\subsubsection{Nuclear Feature Extraction}
We segmented nuclei using CellViT \cite{horst2024cellvit} and extracted morphological, intensity, and texture features using HistomicsTK \cite{pourakpour2025histomicstk}. To ensure robustness against staining variations, all patches underwent deconvolution-based stain normalization prior to feature computation. We extracted a range features per nucleus, including: 
\begin{enumerate}
\item Morphology: Area, perimeter, solidity, and eccentricity.
\item Texture: Haralick features and gradient magnitude histograms.
\item Fractal Dimension: Box-counting dimension to quantify nuclear boundary complexity.
\item Perinuclear Context: Intensity gradients in the immediate perinuclear neighborhood (8-pixel box) to capture local contrast. 
\end{enumerate}

(Mathematical definitions are provided in Supplementary Methods).

\subsubsection{Collagen Architecture Analysis}
To obtain second-harmonic generation (SHG) surrogates for stroma patches, we applied the pre-trained H\&E$\rightarrow$SHG generative model to predict SHG images from H\&E inputs \cite{keikhosravi2020non}. For each synthesized SHG tile, collagen fiber centerlines were extracted using the model’s built-in centerline module, yielding a fiber map. From these maps, we computed 29 features quantifying ECM organization, including Collagen Volume Fraction (ratio of fiber pixels), Fiber Alignment (orientation consistency via structure tensor analysis), and Fiber Geometry (length, thickness, straightness).

\subsubsection{Statistical Analysis (Mixed Effects Modeling)}
To compare RS-positive vs. RS-negative patches with appropriate within-slide control, we sampled 50 predicted positive and 50 predicted negative stroma tiles per patient from the held-out UCLA test cohort (n = 64 patients). 

For each feature, $x$, we fit a logistic generalized linear model (GLM) with slide fixed effects:

\begin{equation}y \sim \text{Bernoulli}(\pi_{i,s}), \space logit(\pi_{i,s}) = \beta_0 + \beta_1 T(x_{i,s}) + \alpha_s\end{equation}

where $y_{i,s} \in \lbrace0, 1\rbrace$ indicates RS-positive for tile $i$ on slide $s$, $\alpha_s$ is a slide indicator (fixed effect) and T($\cdot$) is a per-feature transform chosen by distributional diagnostics (Yeo-Johnson for skewed features; otherwise z-score). Models were estimated in R with glm(). Two-sided p-values were Bonferroni-adjusted across features.

\subsubsection{Handcrafted Benchmark Models}
To benchmark PROTAS against interpretable hand-crafted features, we trained Random Forest, XGBoost, Naive Bayes, and Multi-Layer Perceptron (MLP) classifiers on the extracted nuclear and collagen feature vectors. Models were trained and evaluated on the same patient splits used for PROTAS to ensure a fair head-to-head comparison. Feature importance was quantified using SHAP (SHapley Additive exPlanations) values derived from the best-performing MLP model.

\subsection{Pathologist Survey}
We designed a blinded, external observer study to compare PROTAS against human experts. We randomly sampled 100 stroma-only patches (50 RS-positive, 50 RS-negative) from the held-out UCLA test set. Four board-certified pathologists reviewed the images independently via a custom web interface. For each patch, raters were tasked with (1) Classifying the stroma as "Normal" or "Abnormal"; (2) Reporting their confidence level; and (3) Noting specific visual cues (e.g. nuclear shape, collagen texture) used to make the decision.

We computed accuracy, recall, and precision for each rater using the spatially-defined ground truth labels. To quantify consistency, we calculated Cohen's Kappa ($\kappa$) for all pairwise combinations of raters (inter-rater reliability) and between each rater and the model. (Survey interface screenshots and individual rater metrics in Supplementary Figure S3 and Supplementary Table S10).

\subsection{Transcriptomic Analysis}
\subsubsection{Bulk RNA-Seq (TCGA-PRAD)}
We analyzed 317 primary tumor samples from the TCGA-PRAD cohort. Patients were dichotomized into "RS-rich" (n = 89), and "RS-poor" (n = 228) groups based on the 75th percentile of their PROTAS slide-level scores. Differential gene expression analysis was performed using DESeq2 \cite{love2014moderated} on raw count matrices. Gene Set Enrichment Analysis (GSEA) was conducted using Enrichr \cite{kuleshov2016enrichr} to identify upregulated biological pathways (FDR $<$ 0.05).

\subsubsection{Spatial Transcriptomics (Visium):}
We utilized prostate Visium samples from the HEST-1K dataset \cite{jaume2024hest}. We selected four representative slides with balanced RS distributions. The co-registered H\&E image at each Visium spot was processed by PROTAS to generate a spot-level RS probability. To identify differentially expressed genes while accounting for spatial autocorrelation, we utilized the SPADE framework \cite{qin2024spatial}. We modeled gene expression using a spatial Gaussian process with an RBF kernel optimized for each sample. Within-sample $p$-values were adjusted using the Benjamini-Hochberg procedure, and results were meta-analyzed across the four samples using Fisher's method. (Detailed gene filtering and robustness checks via label permutation are provided in Supplementary Methods).

\subsection{Domain Adaptation for Biopsy Samples}
\subsubsection{Diagnosing Domain Shift}
To assess the compatibility of biopsy and resection data, we randomly sampled 500,000 feature embeddings (119,606 RP; 380,394 biopsy) and visualized their distribution using Uniform Manifold Approximation and Projection (UMAP). The distinct clustering of biopsy versus RP patches confirmed a significant domain mismatch precluding direct model transfer. (Supplementary Figure \ref{supp_fig4:1}).

\subsubsection{Adversarial Training Strategy}
To align these distributions, we adopted a Domain-Adversarial Neural Network (DANN) approach \cite{ganin2016domain}. We augmented the PROTAS architecture with a domain classifier branch trained to distinguish between RP and biopsy patches. This branch was connected via a Gradient Reversal Layer (GRL), which inverts the gradient during backpropagation. This adversarial setup optimizes the feature encoder to minimize the label prediction loss (identifying RS) while simultaneously maximizing the domain classification loss (making RP and biopsy patches indistinguishable).

\subsubsection{Biopsy Ground Truth Definition}
Due to the absence of distant normal tissue is biopsies, we defined labels based on tumor burden.
\begin{enumerate}
\item RS-Positive: Cores containing high-grade cancer (Gleason pattern 4-5) with $\geq$ 50\% or $\geq$ 70\% tumor involvement.
\item RS-Negative: Cores from patients with entirely benign biopsies.

\end{enumerate}

\section{Funding}
This work was supported by the National Cancer Institute of the National Institutes of Health under Award Number R01CA279666 and the National Library of Medicine of the National Institutes of Health under Award Number T15LM013976.

\printbibliography
\clearpage
\textbf{Main Paper Figures}

Figure List:
\begin{itemize}
    \item {Figure 1: Main pipeline figure}
    \item {Figure 2: Slide-level Features and Heatmaps}
    \item {Figure 3: Survival Models}
    \item {Figure 4: Hand-crafted Features}
    \item {Figure 5.1: Volcano Plots of TCGA-PRAD DGE}
    \item {Figure 5.2: Dot Plots from Gene Set Enrichment Analysis}
    \item {Figure 6: Biopsy Dataset and Adversarial Training}
\end{itemize}

\begin{figure}[p]
\vspace*{-2cm}
 \hspace*{-1.5cm}
    \centering
    \includegraphics[width=1.15\textwidth]{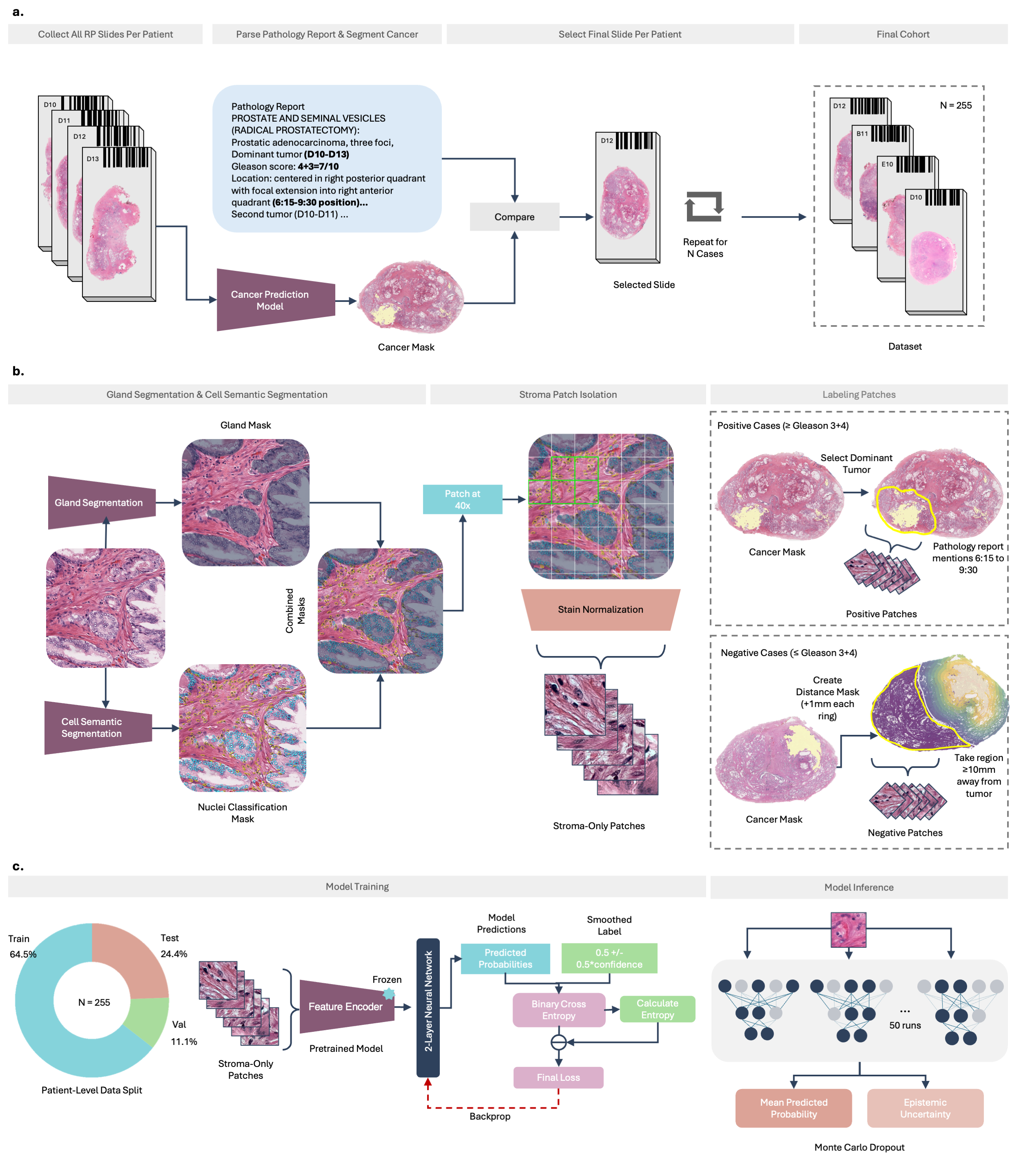}
\end{figure}

\clearpage
\noindent\captionof{figure}{\textbf{Study Pipeline} (a). Cohort curation and slide selection. For each RP case, pathology reports were parsed to identify the dominant tumor focus and location. An automated cancer-prediction model produced tumor masks, which were cross-checked against the report to select one representative whole-mount H\&E WSI per patient, resulting in a final cohort of $N = 255$ (b). Stroma-patch isolation and labeling. Gland segmentation (internal model) and cell semantic segmentation (CellViT) were used to remove glandular epithelium and isolate stroma-only patches. Slides were tiled at 40$\times$ into 256$\times$256 pixel patches, then stain-normalized. Positive stroma patches were sampled from within the dominant tumor region (report-concordant), whereas negative stroma patches were sampled from $\geq$ 10 mm away from any tumor focus. (c). Model training and inference. Patients were split at the patient level (train/validation/test: 64.5\%/11.1\%/24.4\%). A pretrained pathology feature encoder generated fixed embeddings for stroma patches; a two-layer neural network (with batch normalization and dropout) was trained with severity-aware soft labels and an entropy (confidence-penalty) term. At inference, Mont Carlo Dropout provided mean RS probability and epistemic uncertainty for each RS prediction.}
\label{fig:1}

\begin{figure}[p]
\vspace*{-2cm}
 \hspace*{-1.5cm}
    \centering
    \includegraphics[width=1.1\textwidth]{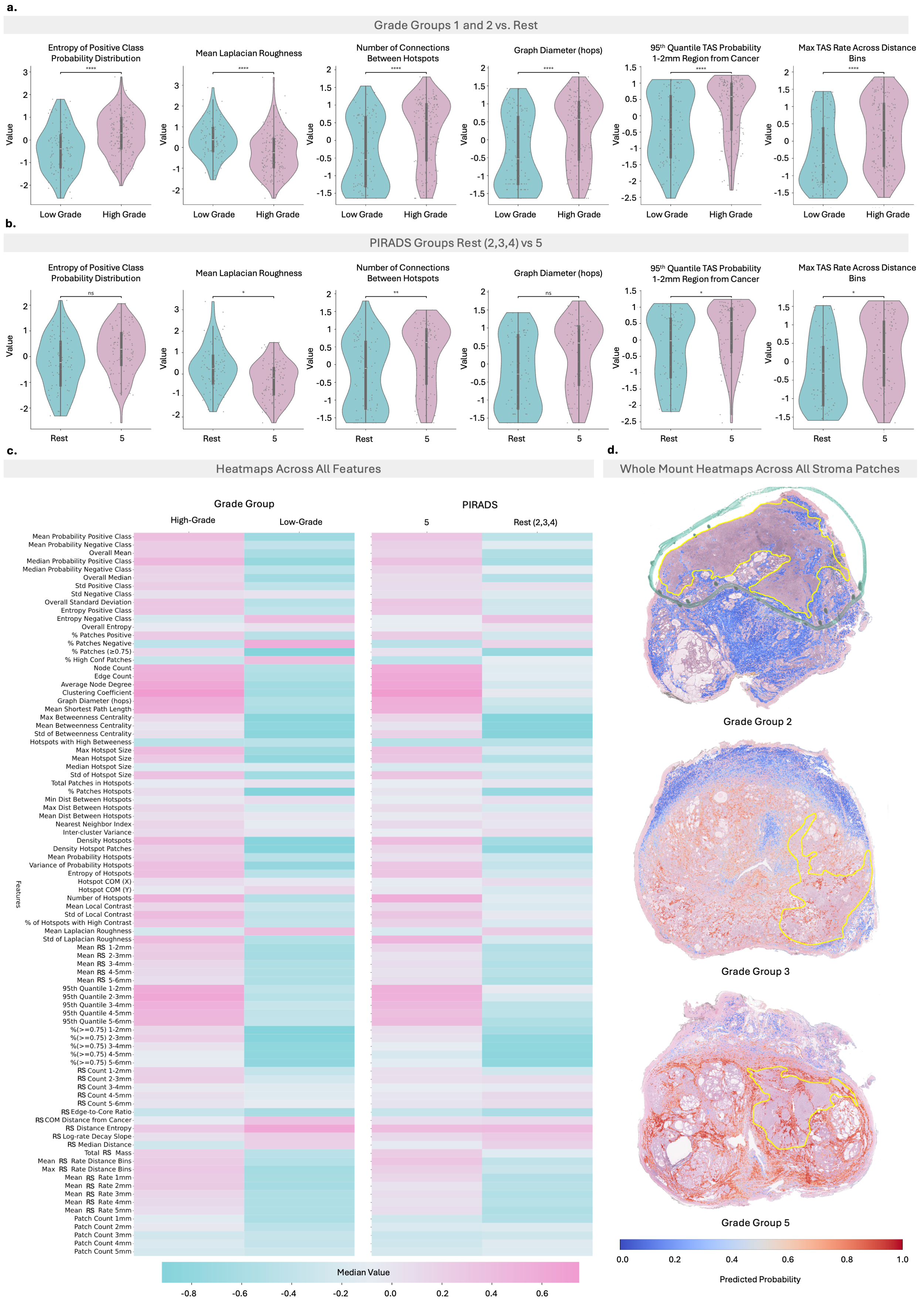}
\end{figure}

\clearpage
\noindent\captionof{figure}{\textbf{Slide-level RS features and spatial maps.} (a). Distributions of six representative slide-level RS features comparing low-grade ($\leq$GG2) versus high-grade ($\geq$GG3): entropy of RS-positive probabilities, hotspot Laplacian mean, number of hotspot-hotspot connections, graph diameter (hops), 95th-percentile RS probability within the 1-2 mm ring outside tumor, and maximum RS rate across 1-mm distance bins. Group differences were assessed with Wilcoxon rank-sum tests with Bonferroni adjustment. (b). The same six features stratified by PI-RADS score (2-4 vs. 5). (c). Heatmap of per-group medians for all slide-level features (left: $\leq$GG2 vs. $\geq$GG3; right: PI-RADS 2-4 vs. 5); deeper magenta indicates a higher median value. (d). Example whole-mount probability maps over stroma-only patches showing PROTAS predictions; warmer colors denote higher RS probability (scale 0-1). Yellow outlines indicate tumor regions.}
\label{fig:2}

\begin{figure}[p]
\vspace*{-2cm}
 \hspace*{-1.5cm}
    \centering
    \includegraphics[width=1.1\textwidth]{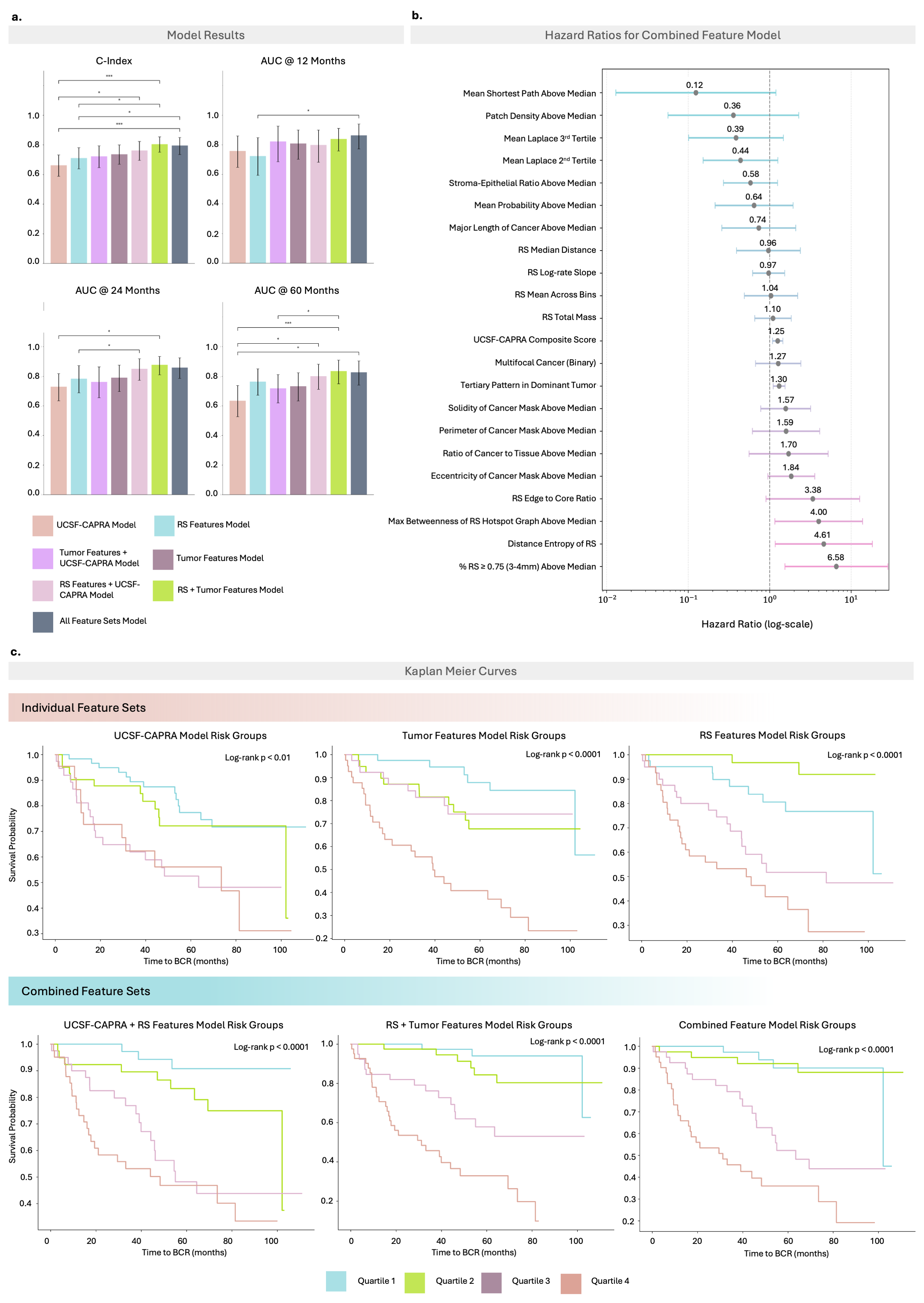}
\end{figure}

\clearpage
\noindent\captionof{figure}{\textbf{Prognostic Modeling with RS features, tumor features, and UCSF-CAPRA} (a). Model comparison across four metrics. Bars show point estimates with 95\% bootstrap CIs (5,000 replicates) for concordance index and time-dependent area under receiver operating characteristic (AUROC) at 12, 24, and 60 months. Asterisks denote significant pairwise differences (bootstrap $p<0.05$) across models. (b). Multivariable Cox model including all feature sets (RS features, tumor features, UCSF-CAPRA): hazard ratios with 95\% CIs on a logarithmic scale; the dashed line indicates HR = 1. Feature definitions and binning/standardization follow Methods. (c). Kaplan-Meier curves stratified by quartiles of each model's partial hazard scores; Adjusted Log-rank $p$-values are reported on each panel.}
\label{fig:3}

\begin{figure}[p]
\vspace*{-2cm}
 \hspace*{-1cm}
    \centering
    \includegraphics[width=1.1\textwidth]{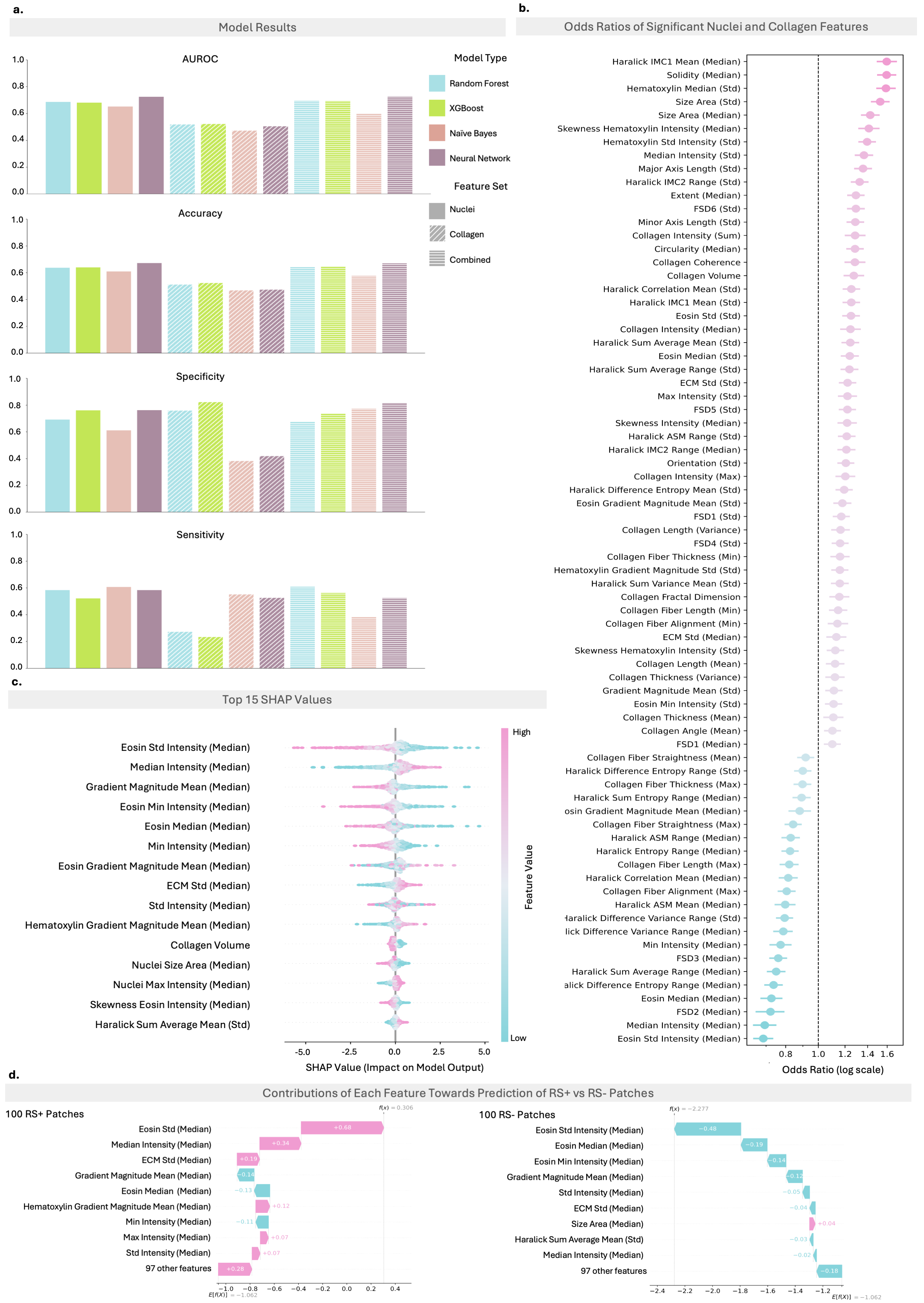}
\end{figure}

\clearpage
\noindent\captionof{figure}{\textbf{Handcrafted feature models and interpretability.} (a). Test set performance of four classifiers (Random Forest, XGBoost, naive Bayes, two-layer neural network) trained on nuclear-only, collagen-only, and combined feature sets. Bars show AUROC, accuracy, specificity, and sensitivity; colors denote model type and hatching denotes feature set. (b). Odds ratios (95\% CIs; log-scale) from a logistic GLM with slide fixed effects for nuclear and collagen features that remained significant after Bonferroni correction; the dashed line marks OR = 1. (c). Top-15 features ranked by mean absolute SHAP value for the MLP trained on combined features. Points show per-patch SHAP values (impact on model); color encodes the feature value (pink = high, blue = low). (d). SHAP waterfall plots for representative RS-positive and RS-negative patches, illustrating how individual feature values shift the prediction from model baseline $E[f(X)]$ to the final output (positive contributions increase RS probability).}
\label{fig:4}

\begin{figure}[p]
 \vspace*{-4cm}
 \hspace*{-1.5cm}
    \centering
    \includegraphics[width=1.15\textwidth]{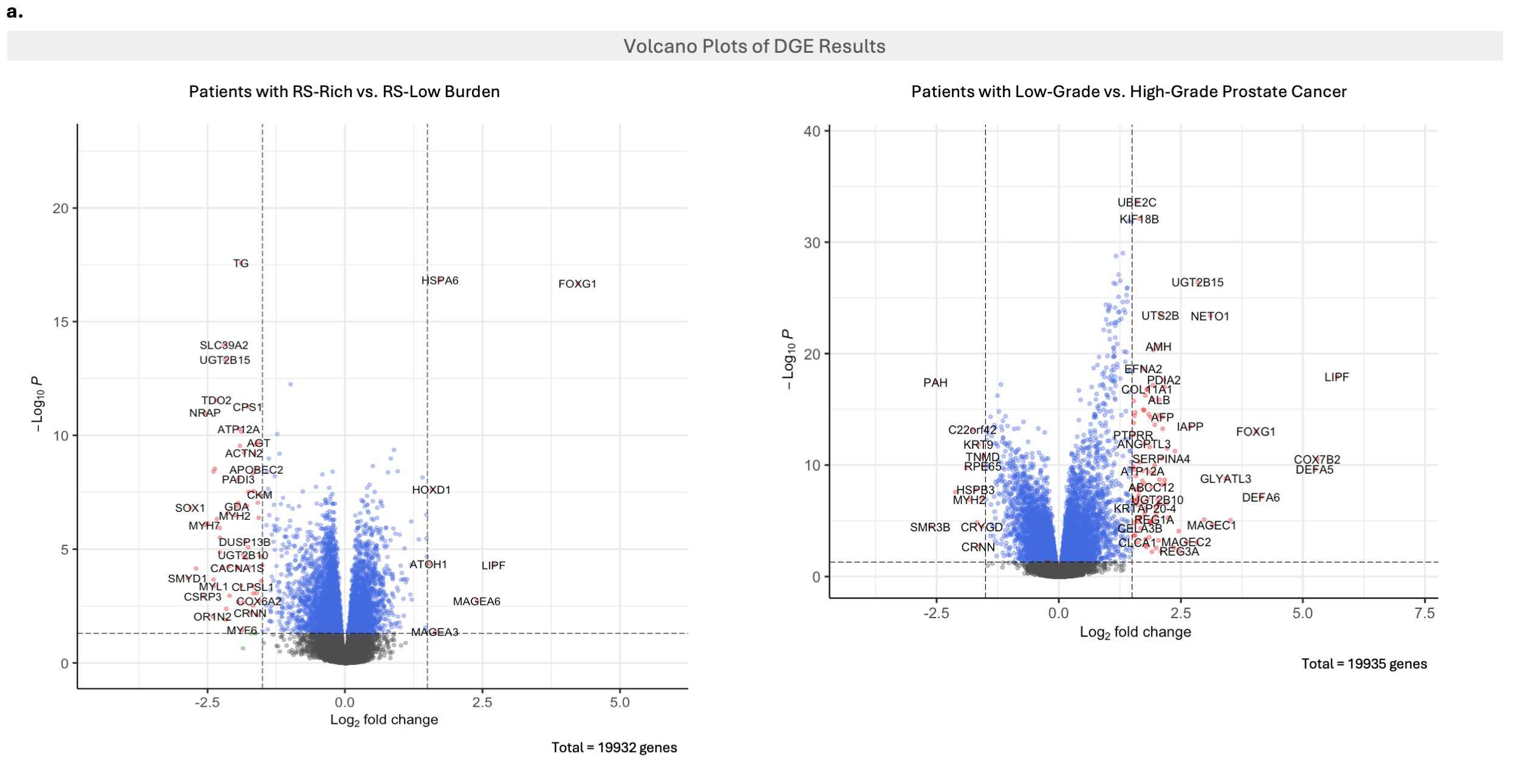}
        \caption{\textbf{TCGA-PRAD differential gene expression} (a) Volcano plots (left) RS-rich vs RS-lower patients (RS-rich defined as the top quartile of the patient-level RS score: proportion of positive patches $\geq$ 75th percentile) and (right) low-grade ($\leq$GG2) vs. high-grade ($\geq$GG3) patients. Points show $\text{log}_2$ fold change (x-axis) versus $-\text{log}_{10}$ adjusted $p$-value (y-axis; DESeq2 with the Benjamini-Hochberg method). Vertical dashed lines mark $|\text{log}_2 \text{FC}| = 1$; the horizontal dashed line marks FDR = $0.05$.}
    \label{fig:5.1}
\end{figure}

\begin{figure}[p]
 \vspace*{-1cm}
 \hspace*{-1.5cm}
    \centering
    \includegraphics[width=1.15\textwidth]{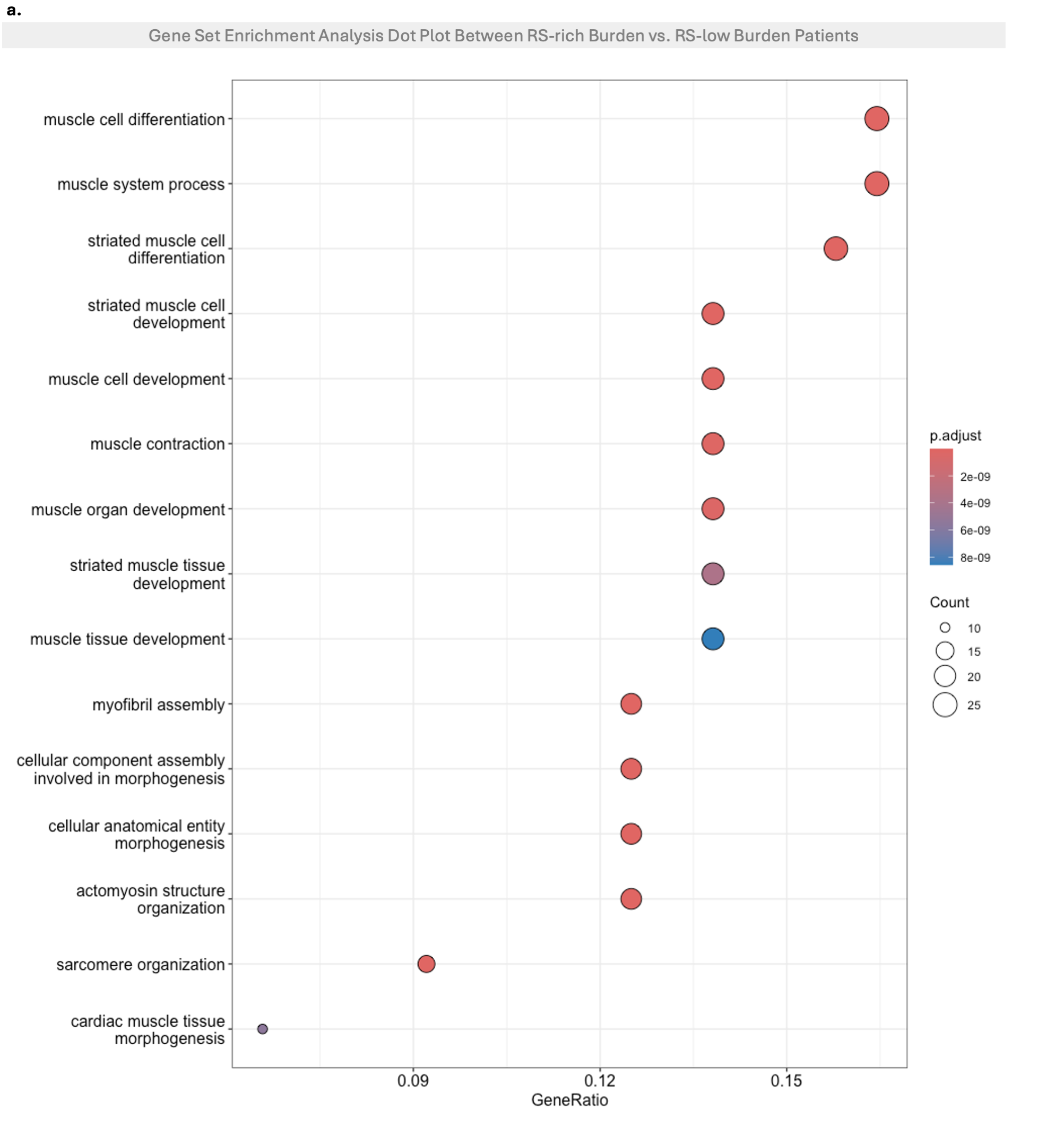}
    \caption{\textbf{Gene set enrichment for RS burden.} Top 15 enriched pathways comparing RS-rich vs. RS-lower patients (RS-rich = top quartile of the patient-level RS score). Points show the GeneRatio on the x-axis; dot size encodes the number of differentially expressed genes in the term; colors denotes Benjamini-Hochberg adjusted $p$-value (EnrichR and DESeq2).}
    \label{fig:5.2.1}
\end{figure}

\begin{figure}[p]
 \vspace*{-1cm}
 \hspace*{-1.5cm}
    \centering
    \includegraphics[width=1.15\textwidth]{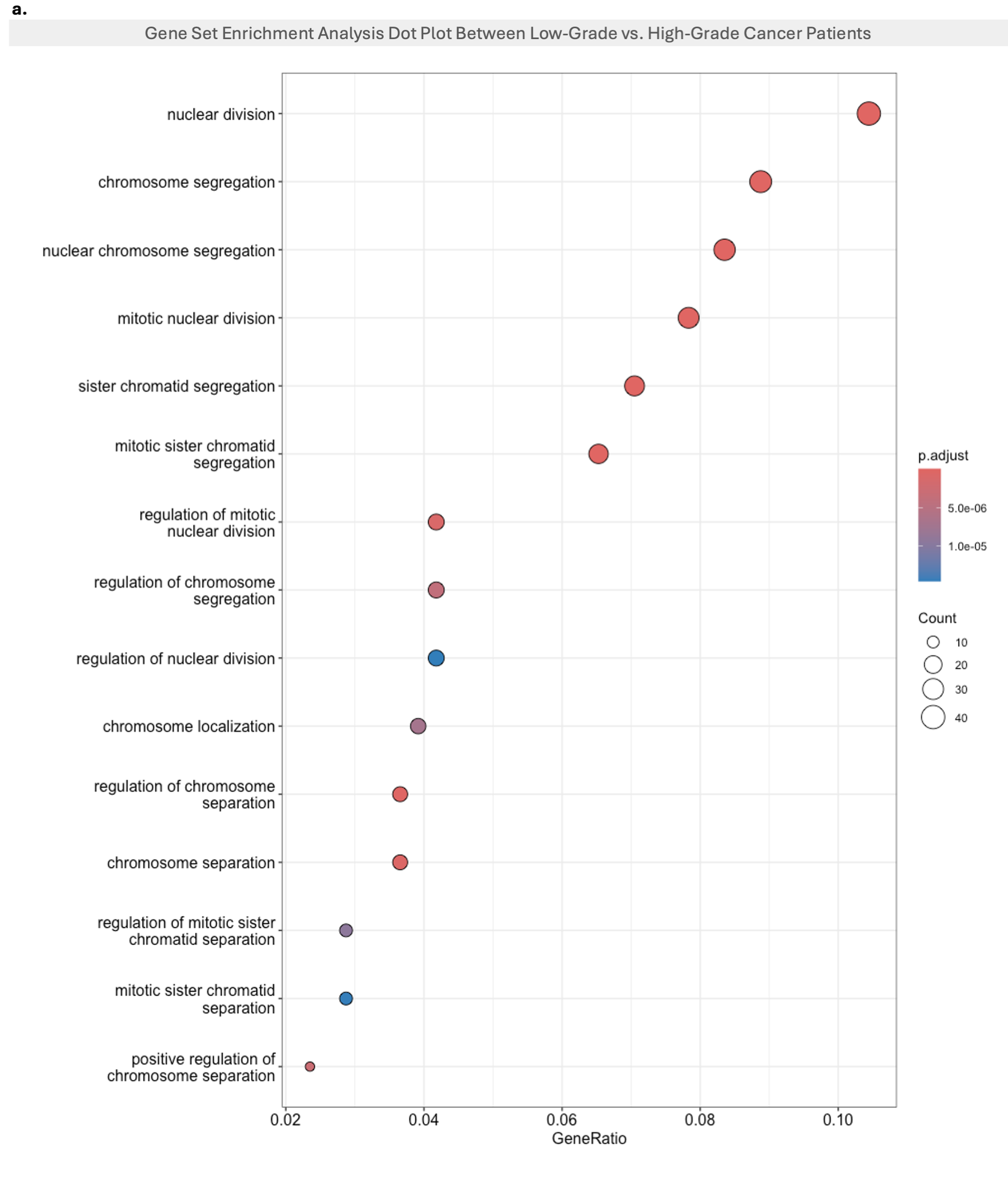}
    \caption{\textbf{Gene set enrichment by grade group.} Top 15 enriched pathways comparing low-grade ($\leq$GG2) vs. high-grade ($\geq$GG3) patients. Points show the GeneRatio on the x-axis; dot size encodes the number of differentially expressed genes in the term; colors denote Benjamini-Hochberg adjusted $p$-value (EnrichR and DESeq2).}
    \label{fig:5.2.2}
\end{figure}

\begin{figure}[p]
 \vspace*{-4cm}
 \hspace*{-1.5cm}
    \centering
    \includegraphics[width=1.15\textwidth]{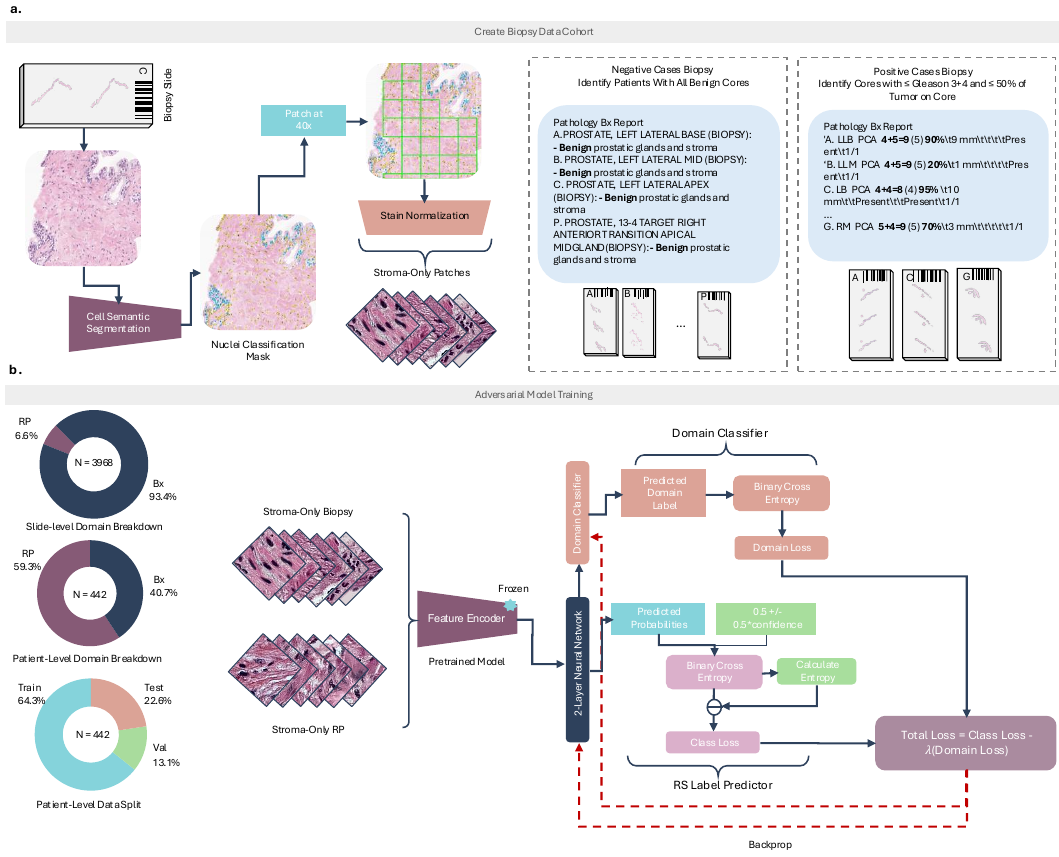}
    \caption{\textbf{Biopsy data processing and domain-adversarial training} (a). Biopsy cohort creation and labeling. For each biopsy WSI, CellViT-based semantic segmentation is used to identify stroma-only patches. Slides are tiled at 40$\times$ magnification, 256$\times$256 pixels, and stain-normalized. Positive labels are assigned to malignant cores based on pathology reports ($\geq$GG3 and $\geq$ 50\% tumor involvement on the core; a $\geq$ 70$\%$ variant was also evaluated, see text). Negative cases were patients whose biopsies were uniformly benign; all cores from these patients were retained for modeling. (b). Domain-adversarial training. Donut plots summarize domain composition and the patient-level train/validation/test split. Stroma-only patches from biopsy and RP were embedded and fed into a two-layer neural network for RS prediction. A lightweight domain classifier (RP vs. Biopsy) attached via a gradient reversal layer encouraged domain-invariant features.}
    \label{fig:6}
\end{figure}

\clearpage
\appendix
\clearpage
\appendix
\renewcommand{\thefigure}{S\arabic{figure}}
\setcounter{figure}{0}

\section{Supplementary Methods}
\subsection{Model Training}
PROTAS was implemented in PyTorch. We utilized the Adam optimizer with a learning rate of $1\times 10^{-4}$ and weight decay of $1 \times 10^{-5}$. Training was conducted with a batch size of 64 for 100 epochs.

\subsubsection{Data Augmentation}
During training 40$\times$ patches underwent random geometric and color augmentations. 

\subsection{Heatmap Generation}
To visualize the distribution of RS over whole-slide images (WSI), we generate a continuous heatmap by spatially aggregating patch-level RS probabilities. Patches extracted at the original magnification (M = 40$\times$, patch size T = 256 pixels) were downsampled by a factor of $D = \frac{M}{m}$ to target a visualization magnification of $m$. The effective patch size on the heatmap grid is:

\begin{equation}T' = \lfloor\frac{T\cdot M}{m\cdot D} \rfloor\end{equation}

\noindent Given N downsampled patch coordinates $\lbrace(x_i,y_i)\rbrace_{i=1}^N$ with RS probabilities $\lbrace p_i\rbrace_{i=1}^N$, we initialize two matrices of size $H_0\times W_0$ where $H_0$ and $W_0$ are the heatmap dimensions.

$$H \in \mathbb{R}^{H_0 \times W_0}$$ 
$$C\in \mathbb{Z}^{H_0 \times W_0}$$

\noindent Where $H$ is the RS probability heatmap, and $C$ is a count map for number of contributions per pixel.

\bigskip

\noindent For each patch $i$, we define the region $\mathcal{R}_i = \lbrace(x,y):x_i\leq x < x_i+T', y_i\leq y < y_i+T' \rbrace$ for all $(x,y) \in \mathcal{R}_i$:

\begin{equation}H[x,y]\leftarrow\frac{H[x,y]\cdot C[x,y]+p_i}{C[x,y]+1}\end{equation}
\begin{equation}C[x,y]\leftarrow C[x,y]+1\end{equation}

\noindent The count map provides a running average to allow for smoothly blending overlapping downsampled patches with each pixel's final value representing the average RS probability across all contributing patches.

\subsection{Slide-Level Quantitative Features}
\subsubsection{Basic Features}
We computed descriptive statistics (mean, median, standard deviation, entropy) of $p_t$ separately for positive ($p_t$ $>$ optimal threshold) and negative predictions, plus the fraction of patches at or above the high-confidence threshold $\tau = 0.75$.

\subsubsection{Hotspot Features}
To characterize spatial clustering of high-confidence RS regions, we identified contiguous RS hotspots and extracted their geometric properties.
\bigskip

\noindent\textit{Grid and Region Definition}

\noindent Patches are mapped to a discrete grid using $x_t^{grid} = \text{int}(x_t/s) - x_{min}$ and $y_t^{grid} = \text{int}(y_t/s) - y_{min}$, creating a binary hotspot grid $B$ where $B[x,y] = 1$ if $p_t \geq \tau$ for the patch at grid location $(x,y)$. Connected components (8-connectivity) define K hotspot regions with sizes $n_k$ and centroids:

$$c_k = s\begin{bmatrix}\bar{x}_k + x_{min} \\ \bar{y}_k + y_{min} \end{bmatrix}$$

where $(\bar{x}_k, \bar{y}_k)$ is the mean grid coordinate of region $k$.

\bigskip
\noindent \textit{Key Spatial Features}
\begin{itemize}
    \item {Region size statistics: $\lbrace \text{min, max, mean, median, std} \rbrace(n_k)$} characterizing hotspot extent
    \item{Inter-region distances: Let $d_{ij} = ||c_i - c_j||_2$ for all pairs. Extract $\lbrace \text{min, max, mean} \rbrace$ of pairwise distances}
    \item{Nearest-neighbor index: \begin{equation}{\text{NNI}} = \frac{\bar{d}_{NN}}{0.5/\sqrt{\rho}}\end{equation} where $\bar{d}_{NN}$ is the mean nearest-neighbor distance and $\rho = K/A$ is regional density over grid area $A$}
    \item{Spatial dispersion: Variance of all pairwise distances $d_{ij}$}
    \item{Density measures: Region density $K/A$ and patch density $N_{hotspot}/A$}
    \item{Local contrast: For each patch $t$ compute contrast $\Delta_t=|p_t-\bar{p}_{neighbors}|$ using the 5-nearest neighbors. Extract mean, standard deviation, and percentage above $0.2$}
    \item{Probability surface roughness: Construct probability grid $T$ and compute Laplacian $\nabla^2T$. Extract mean and standard deviation of $|\nabla^2T|$ at hotspot locations} to quantify local heterogeneity
    \item{Probability-weighted center of mass:
    \begin{equation}x_{COM} = \frac{\sum_{hotspots}p_tx_t}{\sum_{hotspots}p_t}, \quad y_{COM} = \frac{\sum_{hotspots}p_ty_t}{\sum_{hotspots}p_t}\end{equation}}
\end{itemize}

\subsubsection{Graph Hotspot Features}
To capture connectivity patterns among hotspot regions, we constructed a proximity graph $G$ with hotspot centroids $\lbrace c_k\rbrace$ as nodes, connecting regions within distance threshold $\delta = 2000$ pixels. Extracted features include:

\begin{itemize}
    \item {Basic topology: Node count $K$, edge count, mean degree}
    \item{Clustering: Average local clustering coefficient}
    \item{Centrality: Mean, max, standard deviation of betweenness centrality; count above 0.1}
    \item{Path metrics: diameter and average shortest path length (on largest connected component if disconnected)}
\end{itemize}

\subsubsection{Distance-from-Cancer Features}
To characterize RS spatial gradients relative to tumor boundaries, we binned patches by distance $r$ from the cancer mask and computed distance-dependent metrics.

\bigskip

\noindent\textit{Binned Analysis}:
For each distance bin $r$ from cancer boundary, compute:
\begin{itemize}
    \item {Patch count: $n_r$}
    \item {RS sum: $s_r = \sum_{\text{patches in } r} p_t$}
    \item{Mean RS rate: $q_r = \frac{s_r}{n_r}$}
\end{itemize}

\noindent\textit{Aggregate Features}:
Using normalized weights $p_r = \frac{w_r}{\sum_{r}w_r}$ where $w_r = s_r$

\begin{itemize}
    \item {Edge-core ratio: $$\frac{\sum_{r < 1 \text{mm}} w_r}{\sum_{r \geq 3 \text{mm}} w_r}$$}
    \item {Center of mass: $\sum_r d_rp_r$, distance weighted}
    \item {Distribution entropy: $-\sum_r p_r \text{log}(p_r)$}
    \item{Decay rate: slope of $\text{log}(q_r)$ vs distance $d_r$}
    \item {Half-mass distance: distance where cumulative probability reaches 0.5}
    \item{Per-bin outputs: individual $q_r$ and $n_r$ for each distance bin}
\end{itemize}

\subsubsection{Conventional Cancer Features}
For comparison with RS features in prognostic models, we extracted established tumor morphology metrics.

\noindent\textit{Largest Cancer Region Properties}

\bigskip

\noindent For each slide, we identified the largest connected component in the cancer mask (after hole-filling and binary dilation) and extracted standard morphological features using scikit-image regionprops: area, bounding box area, convex area, major/minor axis lengths, eccentricity, equivalent diameter, maximum Feret diameter, perimeter, and solidity \cite{pedregosa2011scikit}.
\bigskip 

\noindent\textit{Area-based Composition Features}

\bigskip

\noindent Cancer-tissue ratios: Using tissue mask $T$ and cancer mask $C$:
\begin{itemize}
    \item {Percent cancer in tissue: $\frac{|C \cap T|}{|T|}$}
    \item {Cancer-tissue ratio: $\frac{|C|}{|T|}$}
\end{itemize}

\subsection{Prognostic Modeling Details}
Selected features were transformed according to distributional diagnostics (skewness, kurtosis, histograms): Yeo-Johnson power transform followed by z-scaling; log transform followed by z-scaling; or z-scaling alone.

For features with large coefficients and large standard errors ($>$ 3) we discretized values into quantile-based bins (two bins or tertiles, depending on the observed distribution).

\subsubsection{Confounder Analysis}
Candidate confounders were: dominant tumor grade, multifocality (binary), percent positive biopsies in preoperative biopsy procedure, age, TNM stage, tertiary pattern in dominant tumor, maximum PSA within 9 months prior to RP procedure, and our computational cancer features.

\subsection{Interpretable Feature Engineering}

\subsubsection{Nuclear Features}
We computed nuclei features using HistomicsTK on CellViT-derived instance masks \cite{pourakpour2025histomicstk, horst2024cellvit}, with stain processing and perinuclear context implemented as follows:

\bigskip
\noindent\textit{Stain normalization} 
Deconvolution-based normalization was applied to transform the input RGB image $I$ using a set target stain matrix $W_{target}$

\begin{equation}I_{normalized} = f(I, W_{target}, W_{estimated})\end{equation}

\noindent Where $W_{estimated}$ is derived using Macenko-PCA unmixing in HistomicsTK package on a randomly selected tissue region for each slide.

\bigskip
\noindent\textit{Color deconvolution}
Stain separation was performed in HistomicsTK:
\begin{equation}S = -\text{log}_{10}(\frac{I_{normalized}}{I_0})\cdot W^{-1}_{complement}\end{equation}

\noindent Where $S$  contains the separated stain channels (hematoxylin, eosin), $I_0$ = 255 is the reference intensity, and $W_{complement}$ is the complemented stain basis matrix.

\bigskip
\noindent\textit{Gradient Magnitude Features}
For each channel $C$ (hematoxylin, eosin, grayscale), gradients are computed as:

\begin{equation}G_x = \frac{\partial C}{\partial x}, \quad G_y = \frac{\partial C}{\partial y}\end{equation}

\begin{equation}|\nabla G| = \sqrt{G_x^2+G_y^2}\end{equation}

\noindent\textit{Intensity Features}
For each channel $C$, we compute:
\begin{itemize}
    \item{Mean: $\mu_C = \frac{1}{N}\sum_{i=1}^N C_i$}
    \item{Median: $\text{median}_C = \text{median}(\lbrace C_i\rbrace)$}
    \item{Standard Deviation: $\sigma_C = \sqrt{\frac{1}{N-1}\sum_{i=1}^N(C_i-\mu_C)^2}$}
    \item{Skewness: $\text{skew}_C = \frac{\frac{1}{N}\sum_{i=1}^N(C_i-\mu_C)^3}{\sigma^3_C}$}
    \item{Interquartile Range: $\text{IQR}_C = Q_{75} - Q_{25}$}
\end{itemize}

\noindent\textit{Gradient Histogram Features}
The gradient magnitude histogram is converted to a probability distribution:

\begin{equation}p_j = \frac{h_j}{\sum_{k=1}^{10} h_k}\end{equation}

\noindent where $h_j$ is the count in bin $j$ of the 10-bin histogram.
From this distribution we then compute:
\begin{itemize}
    \item{Entropy: $H = -\sum_{j=1}^{10}p_j\text{log}_2(p_j)$}
    \item{Energy: $E = \sum_{j=1}^{10}p_j^2$}
\end{itemize}

\noindent\textit{Fractal Dimension}
We use the box counting method to compute the fractal dimension of each nuclei. For nucleus boundary complexity:

\begin{equation}D = -\frac{d(\text{log} N(r))}{d(\text{log}r)}\end{equation}

\noindent where $N(r)$ is the number of boxes of size $r$ needed to cover the nucleus boundary, computed across scales $r = 2^n$ for $n = 1, 2, ..., \text{log}_2(\text{min}(\text{width, height}))$.

\bigskip
\noindent\textit{Nucleus-centric morphology} 
From skimage.measure.regionprops, we extracted:
\begin{itemize}
\item{Basic geometry: area, convex area, perimeter, major/minor axis lengths, equivalent diameter, extent, orientation, eccentricity, solidity, minor/major axis ratio.}

\item{Shape indices:}

\begin{equation}\text{Circularity} = \frac{4\pi\times\text{area}}{\text{perimeter}^2},  \quad \text{Roundness} = \frac{\text{perimeter}/2\pi}{\sqrt{\text{area}/\pi}}\end{equation}

\end{itemize}

\bigskip
\noindent\textit{Haralick and Fourier Shape Descriptor Features}
Refer to HistomicsTK documentation.

\bigskip
\noindent\textit{Feature Aggregation}
For each feature $f$ across all nuclei instances $\lbrace f_1, f_2, ..., f_n\rbrace$ in a tile we compute the mean, median, standard deviation, and range.

\bigskip
\noindent\textit{Perinuclear Context Window}

\medskip
\noindent Spatial Expansion: For each nucleus instance with bounding box $B = [(x_{min},y_{min}), (x_{max},y_{max})]$, perinuclear window $B_{\text{expanded}}$ is:

\begin{equation}B_{\text{expanded}} = [(x_{min}-\delta,y_{min}-\delta), (x_{max}+\delta,y_{max}+\delta)]\end{equation}
\noindent where $\delta = 8$ pixels, with boundary constraints to remain within the tile dimensions.

\medskip
\noindent \textit{Context Feature Computation} All intensity and gradient features are computed on $B_{\text{expanded}}$ with nucleus pixels masked out:
$$C_{\text{context}} = C_{\text{expanded}}\cdot M^c_{\text{nucleus}}$$

\noindent where $M_{\text{nucleus}}^c$ is the complement of the nucleus mask (1 for non-nucleus pixels, 0 for nucleus pixels).

\subsubsection{Collagen Features}

To obtain second-harmonic generation (SHG) surrogates for stroma patches, we applied the pre-trained H\&E$\rightarrow$SHG generative model to predict SHG images from H\&E inputs \cite{keikhosravi2020non}. For each synthesized SHG tile, collagen fiber centerlines were extracted using the model’s built-in centerline module, yielding a fiber map. From these centerlines, we computed patch-level features quantifying fiber angle, orientation, coherence, intensity, alignment, thickness, length, and straightness.

\bigskip
\noindent\textit{Collagen Volume Fraction}
The collagen volume fraction was computed as the ratio of pixels above a threshold intensity:
\begin{equation}\text{Collagen Volume} = \frac{\sum_{i,j}\boldsymbol{1}[I(i,j)>0.05]}{W \times H}\end{equation}

\noindent where $I(i,j)$ is the intensity at pixel $(i,j)$, $\boldsymbol{1}[\cdot]$ is the indicator function, and $W \times H$ represents the total image dimensions, (256 $\times$ 256).

\bigskip
\noindent\textit{Fractal Dimension}
Fractal dimension was estimated using the box-counting method:

\begin{equation}D = -\frac{d(\text{log} N(s))}{d(\text{log}\space s)}\end{equation}

\noindent where $N(s)$ is the number of boxes of size $s$ containing at least one fiber pixel, computed across scales $s = 2^{0.01}, 2^{0.02},...,2^1$.

\bigskip

\noindent\textit{Fiber Orientation Analysis}
Local fiber orientation was quantified using structure tensor analysis. For each pixel, the minimum eigenvector $\boldsymbol{\text{v}}_{min}$ of the structure tensor provided the fiber orientation direction.

\bigskip
\noindent\textit{Angle Deviation}
The deviation between fiber direction and radial direction from image center was computed as:
\begin{equation}\text{Angle Deviation} = \arccos(|d_{norm}\cdot v_{min}|)\times \frac{180}{\pi}\end{equation}

\noindent where $d_{norm}$ is the normalized displacement vector from image center, in our case since our image is 256 x 256 pixels
\begin{equation}d_{norm} = \frac{(x-128)(y-128)}{\sqrt{(x-128)^2+(y-128)^2}}\end{equation}

\noindent\textit{Orientation Deviation}
A normalized orientation measure was calculated as:
\begin{equation}\text{Orientation Deviation}=1-\frac{2 \times \text{Angle Deviation}}{0.5\pi}\end{equation}

\noindent\textit{Coherence Measures}
Fiber coherence was quantified using weighted local orientation:

\begin{equation}\text{Coherence} = \frac{I_{smooth}\times I_{background}}{I_{smooth}}\end{equation}

\noindent Where $I_{smooth}$ is the Gaussian-smoothed SHG image ($\sigma = 0.5$) and $I_{background}$ is the background-smoothed version ($\sigma = 15$).

\bigskip
\noindent\textit{Intensity Features}
Three intensity-based measures were computed:
\begin{itemize}
    \item{Maximum intensity: $I_{max} = \text{max}_{ij} I_{smooth}(i,j)$}
    \item{Mean intensity: $I_{mean} = \frac{1}{W\times H} \sum_{i,j} I_{smooth}(i,j)$} 
    \item{Total intensity: $I_{sum}=\sum_{i,j}I_{smooth}(i,j)$}

\end{itemize}

\bigskip
\noindent\textit{Fiber Geometric Properties}
From extracted centerlines, statistical measures of mean, variance, minimum, and maximum were computed for:
\begin{itemize}
    \item{Alignment: degree of fiber orientation consistency}
    \item{Thickness: average fiber width along centerline}
    \item{Straightness: ratio of end-to-end distance to actual fiber length}
    \item{Angle: fiber orientation angle relative to reference direction}
    \item{Length: total fiber length}
\end{itemize}
\subsection{Genomic Analysis Details}

\subsubsection{TCGA-PRAD Figures}
Dotplots from Figure \ref{fig:5.2.1} and Figure \ref{fig:5.2.2} were generated using the same R package. Volcano plots were generated using the EnhancedVolcano package (Figure \ref{fig:5.1})\cite{https://doi.org/10.18129/b9.bioc.enhancedvolcano}.
\subsection{Visum Preprocessing}
To avoid patient duplication, we retained one Visium slide per patient and further restricted to slides with approximately balanced RS-positive/RS-negative spot counts; slides with high imbalance were excluded, yielding a final set of four Visium samples. To ensure a common gene set across samples, we intersected genes present in all four slides after filtering. Genes were filtered in scanpy to those detected in $\geq$ 5 spots and with $\geq$ 100 total counts; downstream DGE used this filtered intersection \cite{wolf2018scanpy}.

\subsubsection{Meta-analysis and robustness check}
Gene-level statistics were combined across the four samples using Fisher’s method, and the resulting meta-analysis p-values were Benjamini-Hochberg adjusted. As a qualitative robustness check of our DGE approach, we also performed within sample label permutation testing ($B = 10$), shuffling RS labels across spots while preserving group sizes and re-running the pipeline; permutation analyses confirmed our observations exceeded the empirical baseline. 

\subsection{Software and Libraries}
Key software and libraries used for analysis included:
\begin{itemize}
    \item Deep Learning: PyTorch
    \item Survival Analysis: lifelines, scikit-learn
    \item Genomics: DESeq2, Enrichr, Scanpy
    \item Pathology: HistomicsTK, TCGAbiolinks
\end{itemize}

\section{Supplementary Figures and Tables}

\bigskip

\subsection{Figure List:}
\begin{itemize}
    \item {Supplementary Figure \ref{supp_fig1:1}: Deep features vs Nuclei Features Heatmap}
    \item {Supplementary Figure \ref{supp_fig2:1}: Deep features vs Nuclei Features Heatmap}
    \item {Supplementary Figure \ref{supp_fig3:1}: Pathologist Survey App}
    \item {Supplementary Figure \ref{supp_fig4:1}: UMAP of Domain Shift}
\end{itemize}

%\begin{comment}
\subsection{Supplementary Tables}
\bigskip
\begin{itemize}
    \item {Supplementary Table \ref{tab:s1}: Slide-level RS features between $\leq$GG2 vs. $\geq$GG3}
    \item {Supplementary Table \ref{tab:s2}: Slide-level RS features between PIRADS $2-4$ vs. PIRADS $5$}
    \item {Supplementary Table \ref{tab:s3}: Cox PH Model Results with 95\% CI}
    \item {Supplementary Table \ref{tab:s4}: Bootstrap Model Comparison Results}
    \item {Supplementary Table \ref{tab:s5}: Cox PH RS Confounder Results}
    \item {Supplementary Table \ref{tab:s6}: Kaplan-Meier Log-rank Test Results}
    \item {Supplementary Table \ref{tab:s7}: Nuclei Feature GLM Fixed Effect Model Results}
    \item {Supplementary Table \ref{tab:s8}: Collagen Feature GLM Fixed Effect Model Results}
    \item {Supplementary Table \ref{tab:s9}: Hand-crafted Feature Model Results}
    \item {Supplementary Table \ref{tab:s10}: Pathologist Survey Metrics}
    \item {Supplementary Table \ref{tab:s11}: Cohen's $\kappa$ Results}
   
\end{itemize}
%\end{comment}

\setcounter{figure}{0}
\begin{figure}[p]
\vspace*{-2cm}
 \hspace*{-2cm}
    \centering
    \includegraphics[width=1.1\textwidth]{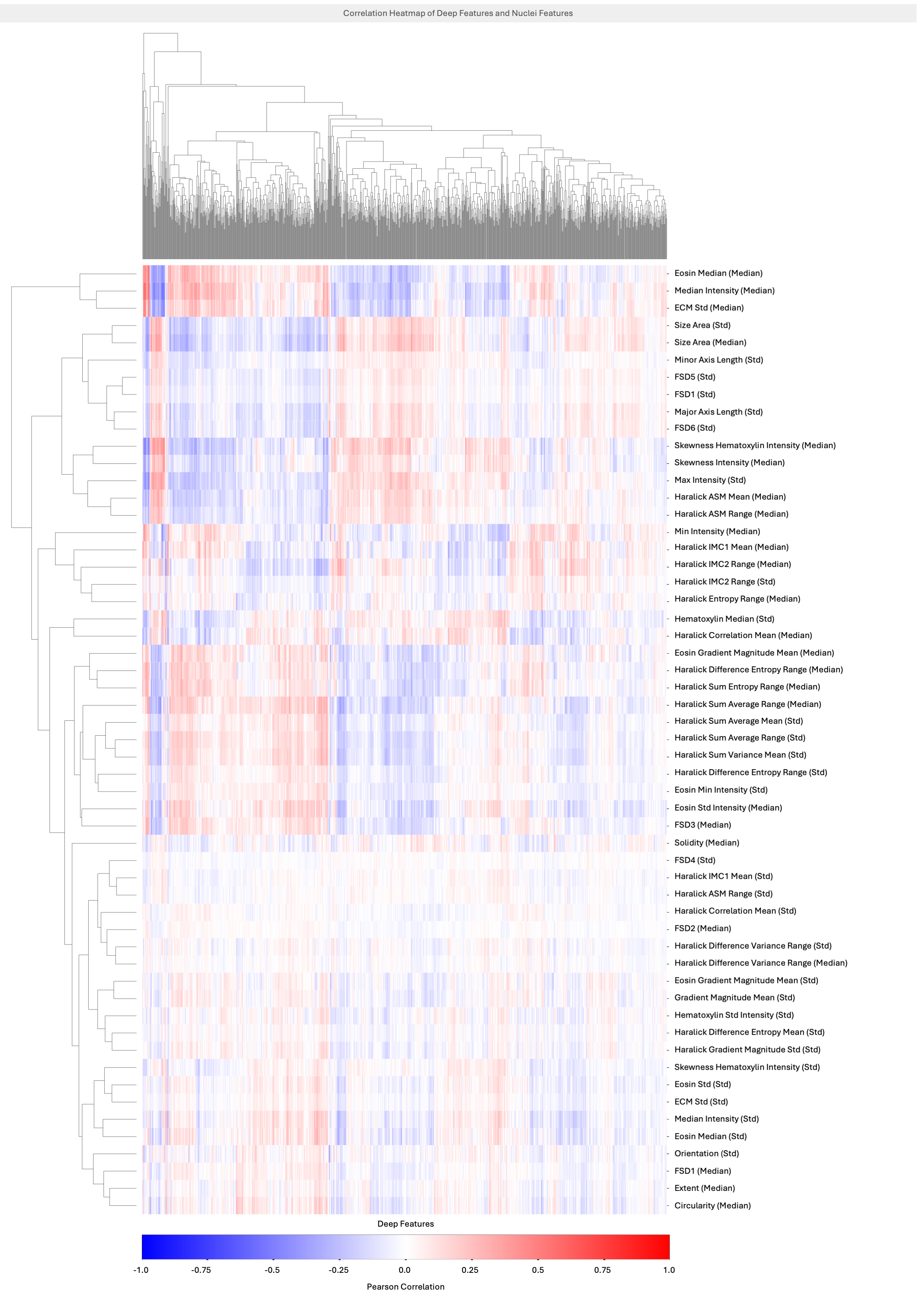}
    
\end{figure}

\clearpage
\noindent\captionof{figure}{Pearson correlation matrix between foundation-model (UNI) embeddings (columns) and hand-crafted nuclei descriptors (rows) aggregated at the 256 x 256 patch level. Rows and columns are hierarchically clustered. Color encodes Pearson $r$. Nuclei features shown are those significant in the fixed-effects logistic GLM (Methods); all variables were normalized prior to correlation.}
\label{supp_fig1:1}

\begin{figure}[p]
\vspace*{-2cm}
 \hspace*{-1cm}
    \centering
    \includegraphics[width=1.1\textwidth]{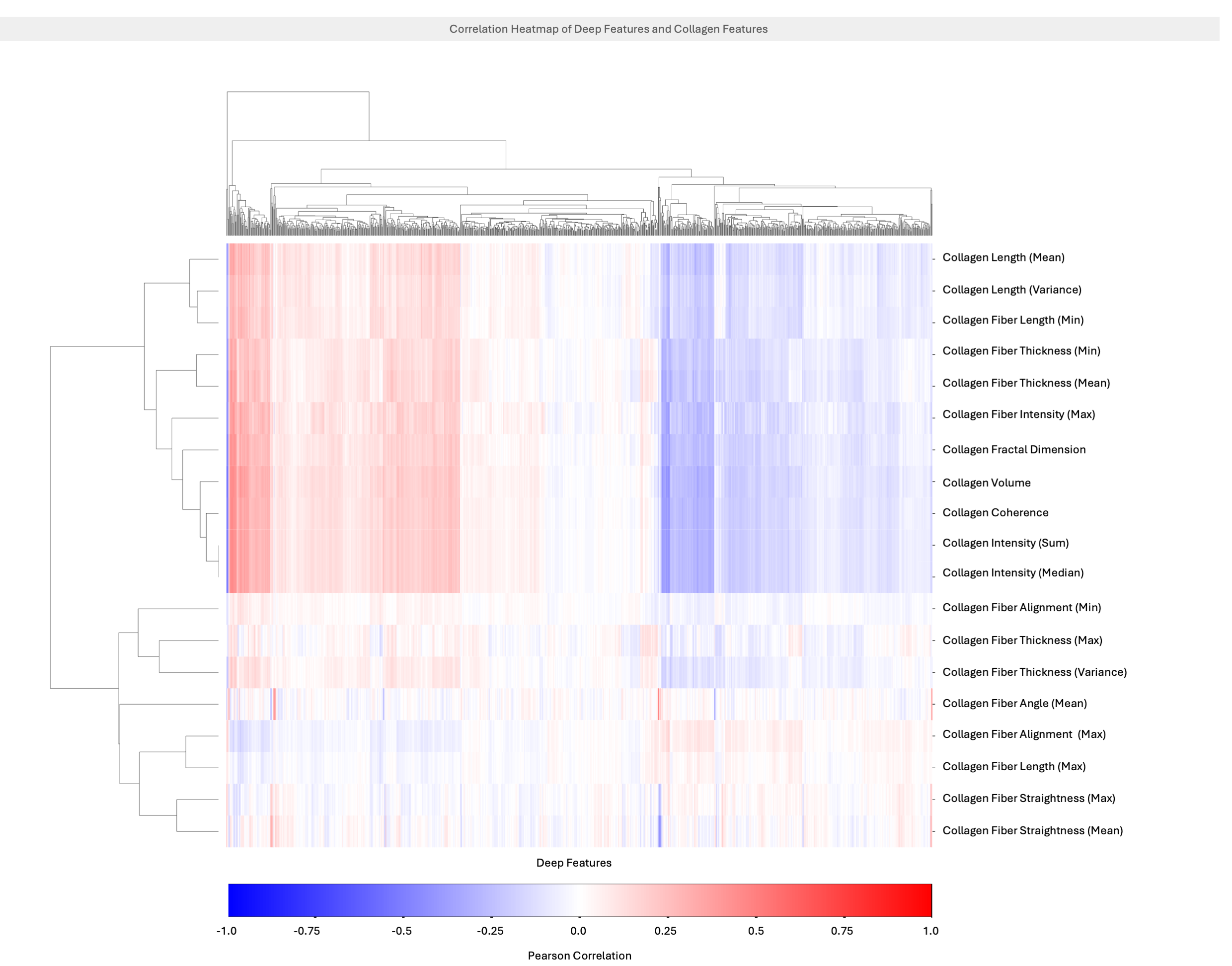}
    \caption{Pearson correlation matrix between foundation-model (UNI) embeddings (columns) and hand-crafted collagen descriptors (rows) aggregated at the 256 x 256 patch level. Rows and columns are hierarchically clustered. Color encodes Pearson $r$. Collagen features shown are those significant in the fixed-effects logistic GLM (Methods); all variables were normalized prior to correlation.}
    \label{supp_fig2:1}
\end{figure}

\begin{figure}[p]
\vspace*{-2cm}
 \hspace*{-1cm}
    \centering
    \includegraphics[width=1.1\textwidth]{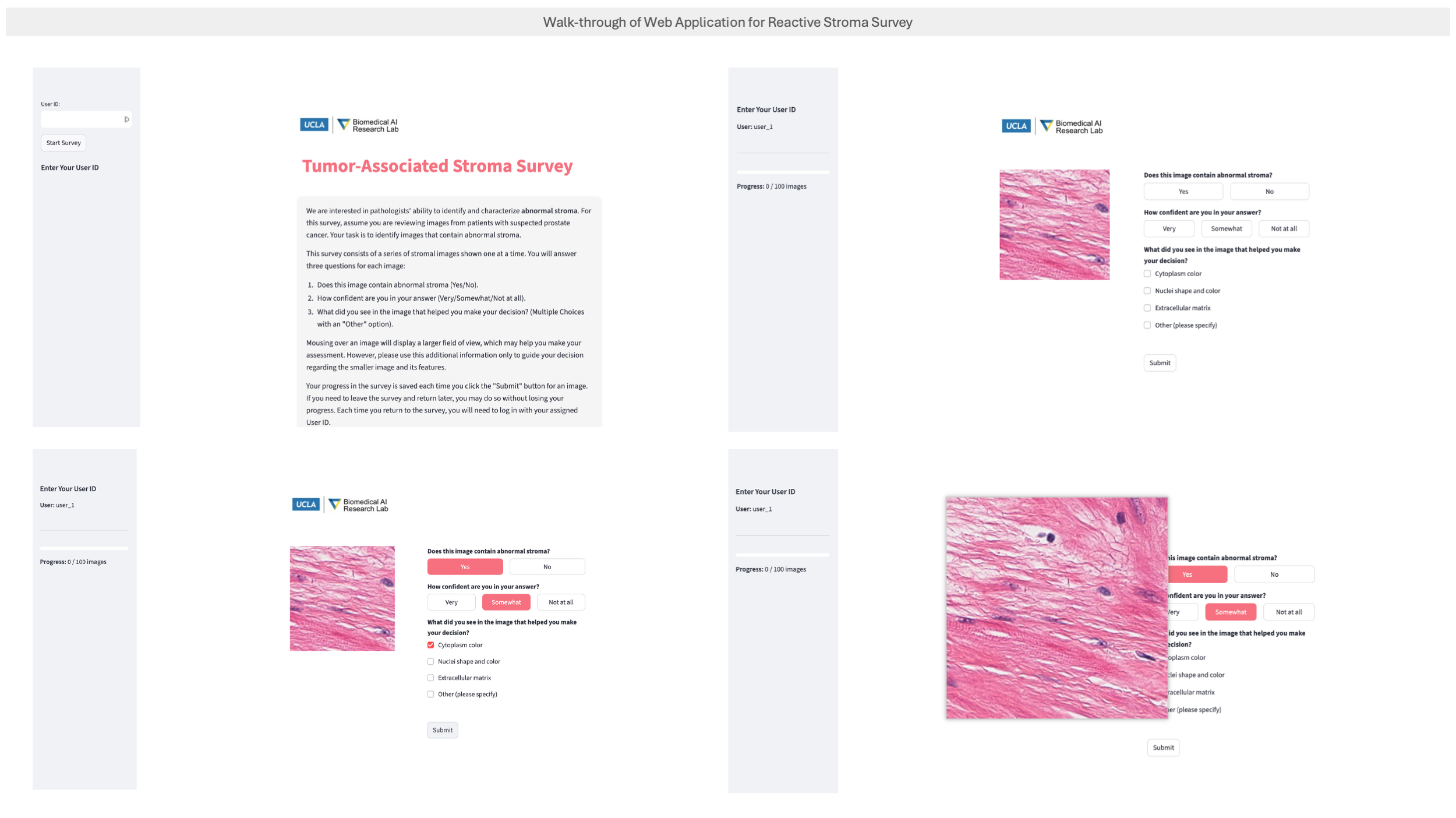}
    \caption{Top-left: landing page with user ID entry and study instructions; Top-right: annotation interface for each stroma-only patch: raters select abnormal vs normal; stroma, record confidence, and optionally note visual cues, progress bar on left tracks out of 100 images; Bottom-left: feedback to rater as they select their responses for each patch; Bottom-right: Hover preview provides a larger field of view to aid raters in decisions}
    \label{supp_fig3:1}
\end{figure}

\begin{figure}[p]
\vspace*{-4cm}
 \hspace*{-1cm}
    \centering
    \includegraphics[width=0.9\textwidth]{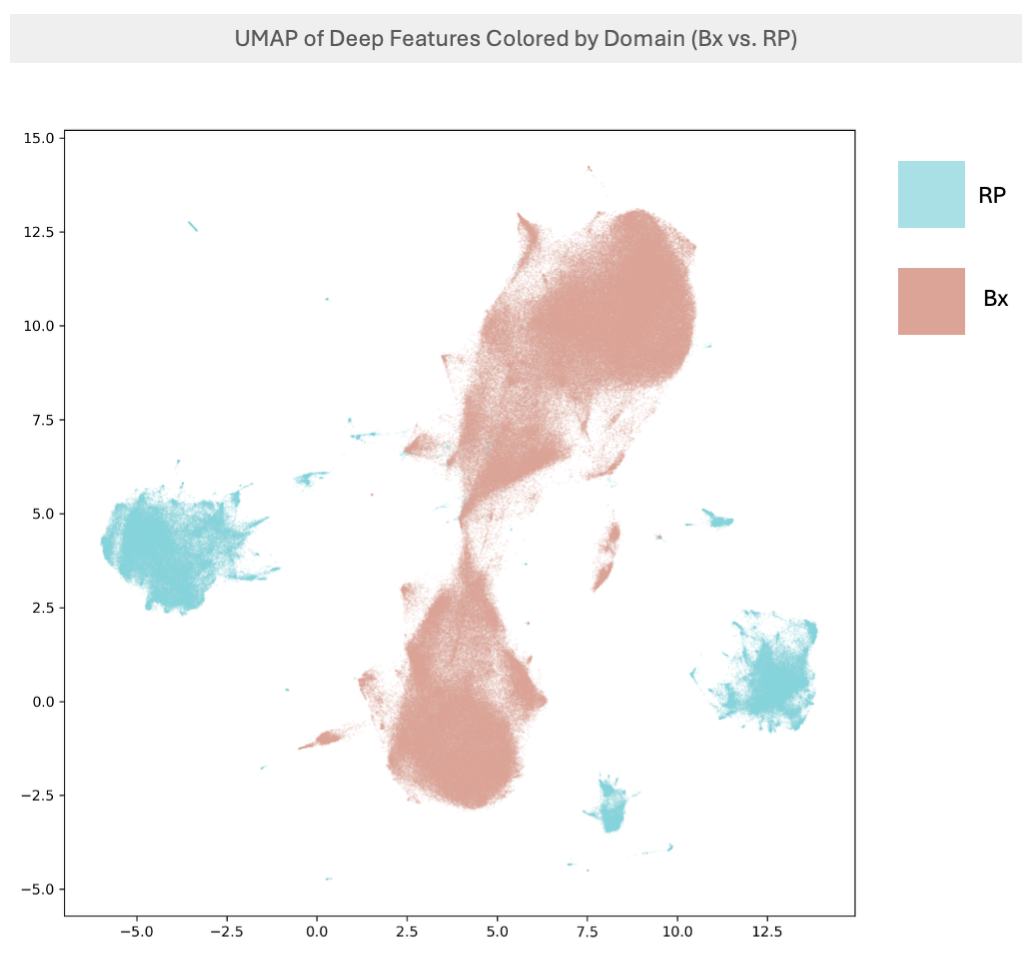}
    \caption{UMAP of patch-level embeddings from the pathology foundation model (UNI), colored by domain (RP = radical prostatectomy, Bx = biopsy). A total of 500,000 patches were randomly sampled from the total training dataset. UMAP was run with $\text{n\_neighbors}$ = 15, $\text{min\_dist}$ = 0.1 using UMAP-learn Python package \cite{mcinnes2018umap}. The clear separation of clusters by domain indicates a distributional mismatch motivating domain-adversarial training adaption (Methods).}
    \label{supp_fig4:1}
\end{figure}
\clearpage

%\begin{comment}
\begin{table}[]
\centering
\begin{tabular}{lrr}
\toprule
\multicolumn{3}{c}{\textbf{Supplementary Table 1 (Panel A)}} \\
\midrule
\multicolumn{1}{c}{\textbf{Feature Name}} & \multicolumn{1}{c}{\textbf{P-value}} & \multicolumn{1}{c}{\textbf{Adjusted P-value}} \\
\midrule
entropy\_pos\_prob & 4.35E-07 & 3.83E-05 \\
median\_prob & 2.01E-06 & 0.00017725 \\
mean\_prob & 2.14E-06 & 0.00018864 \\
RS\_pos\_high\_percent & 2.28E-06 & 0.00020073 \\
RS\_neg\_percent & 3.27E-06 & 0.00028797 \\
mean\_pos\_prob & 3.60E-06 & 0.00031696 \\
median\_pos\_prob & 4.36E-06 & 0.00038355 \\
RS\_pos\_percent & 1.03E-05 & 0.00091049 \\
mean\_neg\_prob & 1.05E-04 & 0.00924896 \\
median\_neg\_prob & 1.52E-04 & 0.0133399 \\
std\_prob & 0.00067527 & 0.05942335 \\
RS\_conf\_percent & 0.00179583 & 0.1580329019 \\
entropy\_neg\_prob & 0.00280078 & 0.2464690727 \\
std\_pos\_prob & 0.01003672477 & 0.8832317801 \\
std\_neg\_prob & 0.7086184909 & 1 \\
entropy\_prob & 0.5456820929 & 1 \\
max\_size & 4.41E-08 & 3.88E-06 \\
std\_size & 6.41E-08 & 5.64E-06 \\
avg\_size & 8.23E-08 & 7.24E-06 \\
patch\_density & 1.46E-07 & 1.28E-05 \\
prob\_variance & 1.62E-07 & 1.42E-05 \\
laplace\_mean & 1.99E-07 & 1.75E-05 \\
region\_density & 4.65E-07 & 4.09E-05 \\
laplace\_std & 6.57E-07 & 5.79E-05 \\
num\_hotspots & 1.14E-06 & 0.00010002 \\
percent\_hotspot & 2.09E-06 & 0.00018367 \\
min\_dist & 7.48E-06 & 0.00065829 \\
entropy\_of\_probs & 1.12E-05 & 0.00098511 \\
std\_contrast & 3.40E-05 & 0.00298967 \\
avg\_prob\_in\_region & 3.59E-05 & 0.00315853 \\
high\_contrast\_percent & 0.00022639 & 0.01992188839 \\
mean\_contrast & 0.00043139 & 0.03796206 \\
spread\_var & 0.6423878322 & 1 \\
median\_size & 0.1153975249 & 1 \\
total\_patches & 0.5468799712 & 1 \\
max\_dist & 0.01303874309 & 1 \\
avg\_dist & 0.1994120553 & 1 \\
nni & 0.6596835925 & 1 \\
center\_of\_mass\_y & 0.884724846 & 1 \\
center\_of\_mass\_x & 0.4954369102 & 1 \\
num\_edges & 3.84E-07 & 3.38E-05 \\
avg\_degree & 4.42E-07 & 3.89E-05 \\
clustering\_coeff & 5.63E-07 & 4.95E-05 \\
\bottomrule
\end{tabular}
\caption{(Panel A) Wilcoxon rank-sum results for each slide-level RS feature comparing patients with $\leq$GG2 vs. $\geq$GG3. P-values were adjusted using Bonferroni correction.}
\label{tab:s1}

\begingroup
    \renewcommand{\thetable}{S\arabic{table}A}
    \label{tab:s1A}
\endgroup
\end{table}

\begin{table}[!htbp]\ContinuedFloat
\centering
\begin{tabular}{lrr}
\toprule
\multicolumn{3}{c}{\textbf{Supplementary Table 1 (Panel B)}} \\
\midrule
\multicolumn{1}{c}{\textbf{Feature Name}} & \multicolumn{1}{c}{\textbf{P-value}} & \multicolumn{1}{c}{\textbf{Adjusted P-value}} \\
\midrule
num\_nodes & 1.14E-06 & 0.00010002 \\
graph\_diameter & 1.72E-06 & 0.00015121 \\
avg\_shortest\_path & 2.79E-06 & 0.00024545 \\
max\_betweenness & 2.12E-05 & 0.00186173 \\
mean\_betweenness & 3.66E-05 & 0.00322092 \\
std\_betweenness & 5.31E-05 & 0.00467523 \\
high\_betweenness\_count & 0.02183311 & 1 \\
q95\_RS\_\_2-3mm & 9.69E-07 & 8.53E-05 \\
q95\_RS\_\_1-2mm & 9.87E-07 & 8.68E-05 \\
burden\_gt\_tau\_\_2-3mm & 2.08E-06 & 0.00018338 \\
RS\_rate\_bin\_2mm & 2.38E-06 & 0.00020982 \\
mean\_RS\_\_2-3mm & 2.38E-06 & 0.00020982 \\
RS\_max\_rate\_over\_bins & 2.38E-06 & 0.00020982 \\
RS\_total\_mass & 3.16E-06 & 0.00027807 \\
burden\_gt\_tau\_\_1-2mm & 3.52E-06 & 0.00031002 \\
mean\_RS\_\_1-2mm & 3.73E-06 & 0.00032818 \\
RS\_rate\_bin\_1mm & 3.73E-06 & 0.00032818 \\
RS\_mean\_rate\_over\_bins & 5.31E-06 & 0.00046743 \\
q95\_RS\_\_3-4mm & 8.40E-06 & 0.00073885 \\
q95\_RS\_\_4-5mm & 9.02E-06 & 0.00079406 \\
RS\_rate\_bin\_3mm & 1.09E-05 & 0.00095702 \\
mean\_RS\_\_3-4mm & 1.09E-05 & 0.00095702 \\
burden\_gt\_tau\_\_3-4mm & 1.39E-05 & 0.00121956 \\
q95\_RS\_\_5-6mm & 1.92E-05 & 0.00168972 \\
RS\_rate\_bin\_4mm & 2.05E-05 & 0.00180695 \\
mean\_RS\_\_4-5mm & 2.05E-05 & 0.00180695 \\
burden\_gt\_tau\_\_4-5mm & 2.18E-05 & 0.00192117 \\
burden\_gt\_tau\_\_5-6mm & 2.46E-05 & 0.00216302 \\
RS\_rate\_bin\_5mm & 5.09E-05 & 0.00447765 \\
mean\_RS\_\_5-6mm & 5.09E-05 & 0.00447765 \\
patch\_count\_bin\_1mm & 0.00015325 & 0.01348563 \\
patch\_count\_\_1-2mm & 0.00015325 & 0.01348563 \\
RS\_edge\_core\_ratio & 0.00220719 & 0.1942325752 \\
patch\_count\_bin\_2mm & 0.00703401 & 0.6189925604 \\
patch\_count\_\_2-3mm & 0.00703401 & 0.6189925604 \\
RS\_com\_distance & 0.00830533 & 0.7308694126 \\
patch\_count\_bin\_3mm & 0.3462258858 & 1 \\
RS\_half\_distance & 0.01429394306 & 1 \\
RS\_lograte\_slope & 0.6754078393 & 1 \\
RS\_distance\_entropy & 0.04358651599 & 1 \\
patch\_count\_\_4-5mm & 0.6415331727 & 1 \\
patch\_count\_\_3-4mm & 0.3462258858 & 1 \\
patch\_count\_bin\_4mm & 0.6415331727 & 1 \\
patch\_count\_\_5-6mm & 0.7537456054 & 1 \\
patch\_count\_bin\_5mm & 0.7537456054 & 1 \\
\bottomrule
\end{tabular}
\caption{(Panel B) Wilcoxon rank-sum results continued for each slide-level RS feature comparing patients with $\leq$GG2 vs. $\geq$GG3. P-values were adjusted using Bonferroni correction.}
\begingroup
    \renewcommand{\thetable}{S\arabic{table}B}
    \label{tab:s1B}
\endgroup
    
\end{table}

\begin{table}[]
\centering
\begin{tabular}{lrr}
\toprule
\multicolumn{3}{c}{\textbf{Supplementary Table 2 (Panel A)}} \\
\midrule
\multicolumn{1}{c}{\textbf{Feature Name}} & \multicolumn{1}{c}{\textbf{P-value}} & \multicolumn{1}{c}{\textbf{Adjusted P-value}} \\
\midrule
RS\_pos\_percent & 4.22E-04 & 3.72E-02 \\
RS\_pos\_high\_percent & 6.40E-04 & 0.05634158 \\
mean\_prob & 6.63E-04 & 0.05835364 \\
median\_prob & 7.19E-04 & 0.0633111 \\
mean\_pos\_prob & 7.71E-04 & 0.06786891 \\
entropy\_pos\_prob & 7.98E-04 & 0.07026034 \\
median\_pos\_prob & 8.96E-04 & 0.07880825 \\
RS\_neg\_percent & 1.71E-03 & 0.15052069 \\
mean\_neg\_prob & 2.37E-02 & 1 \\
median\_neg\_prob & 2.73E-02 & 1 \\
std\_pos\_prob & 0.52964196 & 1 \\
RS\_conf\_percent & 0.12963426 & 1 \\
entropy\_prob & 0.57659929 & 1 \\
entropy\_neg\_prob & 0.01340845 & 1 \\
std\_prob & 0.04754905 & 1 \\
std\_neg\_prob & 0.34986436 & 1 \\
min\_dist & 5.05E-05 & 4.45E-03 \\
laplace\_mean & 1.15E-04 & 1.01E-02 \\
patch\_density & 1.75E-04 & 1.54E-02 \\
laplace\_std & 2.02E-04 & 1.77E-02 \\
region\_density & 2.17E-04 & 1.91E-02 \\
avg\_size & 2.77E-04 & 2.44E-02 \\
num\_hotspots & 2.98E-04 & 2.63E-02 \\
prob\_variance & 3.59E-04 & 3.16E-02 \\
max\_size & 4.44E-04 & 0.03906268083 \\
std\_size & 5.30E-04 & 0.04667526993 \\
percent\_hotspot & 7.03E-04 & 0.06185616176 \\
entropy\_of\_probs & 1.36E-03 & 0.119505889 \\
avg\_prob\_in\_region & 7.08E-03 & 0.6233620241 \\
spread\_var & 5.95E-01 & 1 \\
center\_of\_mass\_x & 0.5679173203 & 1 \\
center\_of\_mass\_y & 0.9872717973 & 1 \\
nni & 0.8858337 & 1 \\
mean\_contrast & 0.1538082216 & 1 \\
std\_contrast & 0.04211049 & 1 \\
high\_contrast\_percent & 0.1187092096 & 1 \\
avg\_dist & 0.9821812417 & 1 \\
max\_dist & 0.2052707431 & 1 \\
total\_patches & 0.5679173203 & 1 \\
median\_size & 0.8368191788 & 1 \\
clustering\_coeff & 3.19E-05 & 2.81E-03 \\
avg\_degree & 4.86E-05 & 4.27E-03 \\
num\_edges & 1.08E-04 & 9.54E-03 \\
\bottomrule
\end{tabular}
\caption{(Panel A) Wilcoxon rank-sum results for each slide-level RS feature comparing PIRADS 2-4 groups vs. PIRADS 5. P-values were adjusted using Bonferroni correction.}
\label{tab:s2}

\begingroup
    \renewcommand{\thetable}{S\arabic{table}A}
    \label{tab:s2A}
\endgroup
\end{table}

\begin{table}[!htbp]\ContinuedFloat
\centering
\begin{tabular}{lrr}
\toprule
\multicolumn{3}{c}{\textbf{Supplementary Table 2 (Panel B)}} \\
\midrule
\multicolumn{1}{c}{\textbf{Feature Name}} & \multicolumn{1}{c}{\textbf{P-value}} & \multicolumn{1}{c}{\textbf{Adjusted P-value}} \\
\midrule
mean\_betweenness & 1.20E-04 & 0.01055066 \\
std\_betweenness & 1.25E-04 & 0.01097066208 \\
max\_betweenness & 1.33E-04 & 0.01170587819 \\
num\_nodes & 2.98E-04 & 0.02626731334 \\
avg\_shortest\_path & 5.22E-04 & 0.04593485 \\
graph\_diameter & 7.89E-04 & 0.06945518 \\
high\_betweenness\_count & 0.02812517 & 1 \\
q95\_RS\_\_1-2mm & 3.31E-04 & 2.92E-02 \\
RS\_max\_rate\_over\_bins & 3.65E-04 & 3.22E-02 \\
mean\_RS\_\_1-2mm & 5.69E-04 & 0.05009127287 \\
RS\_rate\_bin\_1mm & 5.69E-04 & 0.05009127287 \\
burden\_gt\_tau\_\_1-2mm & 5.76E-04 & 0.05066308197 \\
RS\_mean\_rate\_over\_bins & 7.36E-04 & 0.06479779 \\
q95\_RS\_\_2-3mm & 9.16E-04 & 0.08062954 \\
burden\_gt\_tau\_\_3-4mm & 9.26E-04 & 0.08145398 \\
q95\_RS\_\_5-6mm & 1.34E-03 & 0.1175480772 \\
q95\_RS\_\_3-4mm & 1.48E-03 & 0.1305623507 \\
mean\_RS\_\_3-4mm & 1.53E-03 & 0.1349386722 \\
RS\_rate\_bin\_3mm & 1.53E-03 & 0.1349386722 \\
RS\_rate\_bin\_2mm & 1.81E-03 & 0.15891733 \\
mean\_RS\_\_2-3mm & 1.81E-03 & 0.15891733 \\
burden\_gt\_tau\_\_5-6mm & 1.95E-03 & 0.1713590463 \\
burden\_gt\_tau\_\_2-3mm & 2.07E-03 & 0.1825403935 \\
mean\_RS\_\_5-6mm & 2.08E-03 & 0.18327232 \\
RS\_rate\_bin\_5mm & 2.08E-03 & 0.18327232 \\
q95\_RS\_\_4-5mm & 5.26E-03 & 0.4628442338 \\
mean\_RS\_\_4-5mm & 5.53E-03 & 0.4870213518 \\
RS\_rate\_bin\_4mm & 5.53E-03 & 0.4870213518 \\
burden\_gt\_tau\_\_4-5mm & 6.26E-03 & 0.5511084626 \\
RS\_total\_mass & 6.43E-03 & 0.5662101522 \\
RS\_edge\_core\_ratio & 0.00984536 & 0.86639183 \\
patch\_count\_\_1-2mm & 0.01069534946 & 0.9411907527 \\
patch\_count\_bin\_1mm & 0.01069534946 & 0.9411907527 \\
patch\_count\_bin\_3mm & 0.349863747 & 1 \\
patch\_count\_bin\_2mm & 0.7136774982 & 1 \\
patch\_count\_\_4-5mm & 0.2003646332 & 1 \\
patch\_count\_\_5-6mm & 0.1717498509 & 1 \\
RS\_half\_distance & 0.21663342 & 1 \\
RS\_lograte\_slope & 0.4067865365 & 1 \\
RS\_com\_distance & 0.19850557 & 1 \\
RS\_distance\_entropy & 0.6527985636 & 1 \\
patch\_count\_\_3-4mm & 0.349863747 & 1 \\
patch\_count\_bin\_4mm & 0.2003646332 & 1 \\
patch\_count\_\_2-3mm & 0.7136774982 & 1 \\
patch\_count\_bin\_5mm & 0.1717498509 & 1 \\
\bottomrule
\end{tabular}
\caption{(Panel B)  Wilcoxon rank-sum results continued for each slide-level RS feature comparing PIRADS 2-4 groups vs. PIRADS 5.}
\begingroup
    \renewcommand{\thetable}{S\arabic{table}B}
    \label{tab:s2B}
\endgroup
\end{table}

\begin{sidewaystable}[p]
\centering
\small
\begin{tabular}{lcccccccc}
\toprule
\multicolumn{8}{c}{\textbf{Supplementary Table 3}} \\
\midrule
\textbf{Model} & \textbf{C-index} & \textbf{CI} & \textbf{AUC@12} & \textbf{CI} & \textbf{AUC@24} & \textbf{CI} & \textbf{AUC@60} & \textbf{CI} \\
\midrule
\textit{RS Features} & 0.711 & (0.64, 0.78) & 0.724 & (0.59, 0.85) & 0.784 & (0.69, 0.87) & 0.764 & (0.69, 0.85) \\
\textit{RS Features + Tumor Features} & \textbf{0.804} & (0.75, 0.85) & 0.84 & (0.76, 0.91) & \textbf{0.877} & (0.81, 0.93) & \textbf{0.835} & (0.75, 0.91) \\
\textit{RS Features + CAPRA} & 0.762 & (0.7, 0.82) & 0.799 & (0.68, 0.9) & 0.851 & (0.77, 0.92) & 0.8 & (0.71, 0.88) \\
\textit{CAPRA Score} & 0.663 & (0.59, 0.73) & 0.758 & (0.65, 0.86) & 0.73 & (0.63, 0.82) & 0.634 & (0.53, 0.74) \\
\textit{Tumor Features} & 0.736 & (0.67, 0.8) & 0.809 & (0.7, 0.9) & 0.79 & (0.7, 0.87) & 0.732 & (0.63, 0.82) \\
\textit{Combined (all features and CAPRA)} & 0.795 & (0.73, 0.85) & \textbf{0.864} & (0.77, 0.94) & 0.858 & (0.79, 0.92) & 0.827 & (0.74, 0.9) \\
\textit{Tumor Features + CAPRA} & 0.723 & (0.65, 0.79) & 0.823 & (0.69, 0.93) & 0.762 & (0.65, 0.86) & 0.718 & (0.62, 0.81) \\
\bottomrule
\end{tabular}
\caption{Model performance for each Cox Proportional Hazards model reported as mean values from bootstrapped resampling (5,000 replicates) with 95\% confidence intervals. Metrics include concordance index (C-index), and AUC at 12, 24, and 60 months.}
\label{tab:s3}
\end{sidewaystable}

\begin{sidewaystable}[p]
\centering
\scriptsize
\setlength{\tabcolsep}{3pt}
\begin{tabularx}{\textheight}{llXXXX}
\toprule
\multicolumn{6}{c}{\textbf{Supplementary Table 4}} \\
\midrule
\textbf{model1} & \textbf{model2} & \textbf{C-index [95\% CI]} & \textbf{AUC@12 [95\% CI]} & \textbf{AUC@24 [95\% CI]} & \textbf{AUC@60 [95\% CI]} \\
\midrule
\texttt{All\_feature\_preds} & \texttt{Tumor\_plus\_capra\_preds} & -0.072 [-0.139, -0.007] & -0.040 [-0.163, 0.053] & -0.096 [-0.187, -0.017] & -0.107 [-0.196, -0.023] \\
\texttt{Capra\_only\_preds} & \texttt{All\_feature\_preds} & 0.132 [0.060, 0.206] *** & 0.105 [0.017, 0.208] & 0.127 [0.033, 0.226] & 0.191 [0.081, 0.299] * \\
\texttt{Capra\_only\_preds} & \texttt{Tumor\_only\_preds} & 0.074 [0.004, 0.145] & 0.050 [-0.045, 0.146] & 0.060 [-0.034, 0.149] & 0.099 [0.003, 0.196] \\
\texttt{Capra\_only\_preds} & \texttt{Tumor\_plus\_capra\_preds} & 0.060 [-0.010, 0.132] & 0.065 [-0.052, 0.179] & 0.031 [-0.073, 0.131] & 0.084 [-0.007, 0.175] \\
\texttt{Tumor\_only\_preds} & \texttt{All\_feature\_preds} & 0.058 [-0.008, 0.125] & 0.055 [-0.051, 0.163] & 0.067 [-0.015, 0.154] & 0.092 [-0.004, 0.190] \\
\texttt{Tumor\_only\_preds} & \texttt{Tumor\_plus\_capra\_preds} & -0.014 [-0.063, 0.033] & 0.015 [-0.073, 0.093] & -0.029 [-0.106, 0.047] & -0.015 [-0.072, 0.041] \\
\texttt{RS\_only\_preds} & \texttt{All\_feature\_preds} & 0.083 [0.024, 0.148] * & 0.140 [0.046, 0.254] * & 0.074 [-0.005, 0.159] & 0.061 [-0.015, 0.142] \\
\texttt{RS\_only\_preds} & \texttt{Capra\_only\_preds} & -0.048 [-0.150, 0.052] & 0.035 [-0.122, 0.192] & -0.053 [-0.187, 0.075] & -0.130 [-0.266, 0.009] \\
\texttt{RS\_only\_preds} & \texttt{Tumor\_only\_preds} & 0.025 [-0.074, 0.129] & 0.085 [-0.085, 0.264] & 0.007 [-0.129, 0.140] & -0.031 [-0.167, 0.106] \\
\texttt{RS\_only\_preds} & \texttt{Tumor\_plus\_capra\_preds} & 0.011 [-0.092, 0.112] & 0.100 [-0.081, 0.267] & -0.022 [-0.161, 0.115] & -0.046 [-0.179, 0.084] \\
\texttt{RS\_only\_preds} & \texttt{RS\_plus\_capra\_preds} & 0.050 [0.012, 0.091] & 0.075 [0.013, 0.147] & 0.067 [0.023, 0.116] * & 0.036 [-0.016, 0.088] \\
\texttt{RS\_only\_preds} & \texttt{RS\_plus\_Tumor\_preds} & 0.093 [0.031, 0.160] * & 0.117 [0.008, 0.236] & 0.094 [0.012, 0.183] & 0.070 [-0.013, 0.157] \\
\texttt{RS\_plus\_capra\_preds} & \texttt{All\_feature\_preds} & 0.033 [-0.011, 0.079] & 0.065 [0.004, 0.143] & 0.007 [-0.051, 0.067] & 0.026 [-0.035, 0.088] \\
\texttt{RS\_plus\_capra\_preds} & \texttt{Capra\_only\_preds} & -0.099 [-0.174, -0.025] * & -0.040 [-0.155, 0.070] & -0.120 [-0.225, -0.024] & -0.166 [-0.268, -0.064] * \\
\texttt{RS\_plus\_capra\_preds} & \texttt{Tumor\_only\_preds} & -0.025 [-0.112, 0.064] & 0.010 [-0.135, 0.161] & -0.060 [-0.175, 0.050] & -0.067 [-0.185, 0.052] \\
\texttt{RS\_plus\_capra\_preds} & \texttt{Tumor\_plus\_capra\_preds} & -0.039 [-0.127, 0.048] & 0.025 [-0.130, 0.169] & -0.088 [-0.211, 0.026] & -0.082 [-0.195, 0.030] \\
\texttt{RS\_plus\_Tumor\_preds} & \texttt{All\_feature\_preds} & -0.010 [-0.045, 0.022] & 0.023 [-0.037, 0.081] & -0.020 [-0.071, 0.027] & -0.009 [-0.051, 0.031] \\
\texttt{RS\_plus\_Tumor\_preds} & \texttt{Capra\_only\_preds} & -0.142 [-0.215, -0.073] *** & -0.082 [-0.179, 0.005] & -0.147 [-0.247, -0.055] * & -0.200 [-0.304, -0.098] *** \\
\texttt{RS\_plus\_Tumor\_preds} & \texttt{Tumor\_only\_preds} & -0.068 [-0.128, -0.011] & -0.032 [-0.134, 0.057] & -0.087 [-0.166, -0.019] & -0.101 [-0.182, -0.021] \\
\texttt{RS\_plus\_Tumor\_preds} & \texttt{Tumor\_plus\_capra\_preds} & -0.082 [-0.161, -0.009] & -0.017 [-0.162, 0.094] & -0.116 [-0.230, -0.014] & -0.116 [-0.209, -0.030] * \\
\texttt{RS\_plus\_Tumor\_preds} & \texttt{RS\_plus\_capra\_preds} & -0.043 [-0.092, 0.004] & -0.042 [-0.128, 0.038] & -0.027 [-0.091, 0.035] & -0.034 [-0.102, 0.031] \\
\bottomrule
\end{tabularx}
\caption{Bootstrap comparison (5,000 replicates) of model performance across all pairs of models for each evaluation metric. Values represent the difference (model$1$-model$2$) in C-index and AUC at 12, 24, and 60 months, with 95\% confidence intervals.} 
\label{tab:s4}
\end{sidewaystable}

\begin{sidewaystable}[p]
\centering
\scriptsize
\setlength{\tabcolsep}{3pt}      % tighter horizontal spacing
\renewcommand{\arraystretch}{1.1}
% remove extra vertical whitespace at top/bottom of sideways page
\setlength{\rotFPtop}{0pt plus 1fil}
\setlength{\rotFPbot}{0pt plus 1fil}

\begin{tabularx}{\textheight}{>{\raggedright\arraybackslash}p{3.8cm}*{9}{>{\centering\arraybackslash}X}}
\toprule
\multicolumn{10}{c}{\textbf{Supplementary Table 5}} \\
\midrule
\textbf{Covariate} &
\textbf{All Clinical} &
\textbf{Dominant Tumor Grade} &
\textbf{Multifocality} &
\textbf{Tumor Features} &
\textbf{PPB} &
\textbf{Age} &
\textbf{TNM Stage} &
\textbf{Tertiary Gleason Pattern} &
\textbf{PSA} \\
\midrule
Average Probability in Hotspots (High vs Low) &
0.858 (p=0.800, CI [0.26, 2.8]) &
0.966 (p=0.948, CI [0.34, 2.75]) &
1.052 (p=0.920, CI [0.39, 2.82]) &
1.017 (p=0.974, CI [0.37, 2.78]) &
1.096 (p=0.857, CI [0.40, 2.99]) &
1.051 (p=0.921, CI [0.39, 2.81]) &
1.055 (p=0.917, CI [0.38, 2.89]) &
0.809 (p=0.697, CI [0.28, 2.35]) &
1.033 (p=0.949, CI [0.38, 2.78]) \\
Average Shortest Path (High vs Low) &
0.167 (p=0.087, CI [0.02, 1.30]) &
0.101 (p=0.011, CI [0.02, 0.59]) * &
0.132 (p=0.022, CI [0.02, 0.75]) * &
0.143 (p=0.032, CI [0.02, 0.85]) * &
0.136 (p=0.031, CI [0.02, 0.84]) * &
0.131 (p=0.021, CI [0.02, 0.74]) * &
0.162 (p=0.050, CI [0.03, 1.00]) &
0.173 (p=0.054, CI [0.03, 1.03]) &
0.132 (p=0.022, CI [0.02, 0.75]) * \\
Amount of High Grade RS in 3--4mm (High vs Low) &
3.439 (p=0.112, CI [0.75, 15.77]) &
5.366 (p=0.018, CI [1.34, 21.52]) * &
4.401 (p=0.032, CI [1.13, 17.11]) * &
3.841 (p=0.045, CI [1.03, 14.30]) * &
4.877 (p=0.023, CI [1.24, 19.14]) * &
4.397 (p=0.033, CI [1.13, 17.11]) * &
5.150 (p=0.019, CI [1.31, 20.26]) * &
3.865 (p=0.051, CI [1.00, 14.98]) &
4.511 (p=0.030, CI [1.15, 17.63]) * \\
Mean Laplacian Roughness (Med vs High/Low) &
0.544 (p=0.284, CI [0.18, 1.66]) &
0.420 (p=0.089, CI [0.15, 1.14]) &
0.412 (p=0.084, CI [0.15, 1.12]) &
0.476 (p=0.140, CI [0.18, 1.28]) &
0.464 (p=0.113, CI [0.18, 1.20]) &
0.418 (p=0.082, CI [0.16, 1.12]) &
0.422 (p=0.075, CI [0.16, 1.09]) &
0.433 (p=0.108, CI [0.16, 1.20]) &
0.416 (p=0.080, CI [0.16, 1.11]) \\
Mean Laplacian Roughness (High vs Med/Low) &
0.688 (p=0.565, CI [0.19, 2.46]) &
0.423 (p=0.150, CI [0.13, 1.36]) &
0.385 (p=0.118, CI [0.12, 1.27]) &
0.507 (p=0.268, CI [0.15, 1.69]) &
0.440 (p=0.171, CI [0.14, 1.43]) &
0.390 (p=0.117, CI [0.12, 1.27]) &
0.405 (p=0.125, CI [0.13, 1.29]) &
0.573 (p=0.364, CI [0.17, 1.91]) &
0.390 (p=0.116, CI [0.12, 1.26]) \\
Max Betweenness (High vs Low) &
3.504 (p=0.039, CI [1.06, 11.53]) * &
4.188 (p=0.010, CI [1.40, 12.53]) * &
3.003 (p=0.049, CI [1.01, 8.96]) * &
2.327 (p=0.134, CI [0.77, 7.02]) &
2.511 (p=0.111, CI [0.81, 7.78]) &
3.010 (p=0.049, CI [1.01, 9.01]) * &
2.714 (p=0.071, CI [0.92, 8.02]) &
2.498 (p=0.101, CI [0.84, 7.46]) &
2.960 (p=0.051, CI [1.00, 8.80]) \\
Patch Density of Hotspots (High vs Low) &
0.555 (p=0.533, CI [0.09, 3.54]) &
0.467 (p=0.343, CI [0.10, 2.25]) &
0.425 (p=0.261, CI [0.10, 1.89]) &
0.545 (p=0.442, CI [0.12, 2.56]) &
0.456 (p=0.319, CI [0.10, 2.14]) &
0.437 (p=0.273, CI [0.10, 1.92]) &
0.511 (p=0.392, CI [0.11, 2.37]) &
0.566 (p=0.464, CI [0.12, 2.60]) &
0.430 (p=0.255, CI [0.10, 1.84]) \\
RS Distance Entropy (Normalized) &
5.423 (p=0.013, CI [1.43, 20.55]) * &
6.675 (p=0.003, CI [1.90, 23.41]) ** &
5.109 (p=0.015, CI [1.37, 19.06]) * &
3.286 (p=0.088, CI [0.84, 12.87]) &
5.073 (p=0.016, CI [1.36, 18.94]) * &
5.083 (p=0.016, CI [1.36, 19.01]) * &
5.183 (p=0.013, CI [1.41, 19.06]) * &
4.722 (p=0.018, CI [1.31, 17.04]) * &
5.087 (p=0.015, CI [1.37, 18.94]) * \\
RS Edge-to-Core Ratio (Normalized) &
3.596 (p=0.084, CI [0.84, 15.33]) &
5.803 (p=0.010, CI [1.53, 21.96]) ** &
4.174 (p=0.035, CI [1.11, 15.72]) * &
2.500 (p=0.211, CI [0.60, 10.49]) &
4.281 (p=0.031, CI [1.14, 16.05]) * &
4.142 (p=0.038, CI [1.08, 15.82]) * &
4.105 (p=0.033, CI [1.12, 14.99]) * &
2.785 (p=0.123, CI [0.76, 10.25]) &
4.094 (p=0.038, CI [1.08, 15.45]) * \\
RS Median Distance (Normalized) &
0.793 (p=0.608, CI [0.33, 1.92]) &
0.763 (p=0.502, CI [0.35, 1.68]) &
0.760 (p=0.465, CI [0.36, 1.59]) &
0.922 (p=0.839, CI [0.42, 2.01]) &
0.801 (p=0.562, CI [0.38, 1.69]) &
0.755 (p=0.460, CI [0.36, 1.59]) &
0.746 (p=0.444, CI [0.35, 1.58]) &
0.665 (p=0.305, CI [0.31, 1.45]) &
0.754 (p=0.454, CI [0.36, 1.58]) \\
RS Log-rate Decay Slope with Distance (Normalized) &
1.082 (p=0.737, CI [0.68, 1.72]) &
1.266 (p=0.279, CI [0.83, 1.94]) &
1.284 (p=0.242, CI [0.84, 1.95]) &
1.296 (p=0.224, CI [0.85, 1.97]) &
1.318 (p=0.174, CI [0.88, 1.96]) &
1.289 (p=0.228, CI [0.85, 1.95]) &
1.362 (p=0.155, CI [0.89, 2.08]) &
1.008 (p=0.969, CI [0.66, 1.53]) &
1.271 (p=0.264, CI [0.83, 1.93]) \\
Mean RS Rate Across Distance Bins (Normalized) &
0.806 (p=0.564, CI [0.39, 1.68]) &
0.718 (p=0.349, CI [0.36, 1.44]) &
0.908 (p=0.773, CI [0.47, 1.75]) &
0.902 (p=0.760, CI [0.46, 1.75]) &
0.938 (p=0.848, CI [0.49, 1.80]) &
0.906 (p=0.767, CI [0.47, 1.74]) &
0.798 (p=0.504, CI [0.41, 1.55]) &
1.039 (p=0.914, CI [0.52, 2.07]) &
0.911 (p=0.781, CI [0.47, 1.76]) \\
Total RS Mass (Normalized) &
1.261 (p=0.381, CI [0.75, 2.12]) &
1.183 (p=0.464, CI [0.75, 1.86]) &
1.236 (p=0.333, CI [0.80, 1.90]) &
1.410 (p=0.136, CI [0.90, 2.22]) &
1.304 (p=0.231, CI [0.85, 2.01]) &
1.232 (p=0.352, CI [0.79, 1.91]) &
1.174 (p=0.467, CI [0.76, 1.81]) &
1.136 (p=0.576, CI [0.73, 1.78]) &
1.218 (p=0.377, CI [0.79, 1.89]) \\
\bottomrule
\end{tabularx}
\caption{Cox proportional hazards models evaluating associations between RS features and clinical covariates. Hazard ratios (HR), p-value, and 95\% confidence intervals are show. "All clinical" includes all covariates jointly; PPB = percent positive biopsy cores, TNM = pathological TNM stage, PSA = prostate-specific antigen.}
\label{tab:s5}
\end{sidewaystable}

\begin{table}[]
\begin{tabular}{lrrr}
\toprule
\multicolumn{4}{c}{\textbf{Supplementary Table 6}} \\
\midrule
\textbf{Model Name} & \multicolumn{1}{l}{\textbf{Test Stat}} & \multicolumn{1}{l}{\textbf{P-Value}}& \multicolumn{1}{l}{\textbf{Adjusted P-Value}} \\
\midrule
\textit{RS Features} & 37.0293613 & 4.54E-08 & 1.06E-07 \\
\textit{RS + Tumor Features} & 64.4035162 & 6.73E-14 & 4.71E-13 \\
\textit{RS Features + UCSF-CAPRA} & 31.8239388 & 5.70E-07 & 7.98E-07 \\
\textit{UCSF-CAPRA} & 13.4545977 & 0.00374988 & 0.00374988 \\
\textit{Tumor Features} & 34.9456299 & 1.25E-07 2.19E-07 \\
\textit{UCSF-CAPRA + RS + Tumor Features} & 57.5461547 & 1.96E-12 & 6.88E-12 \\
\textit{Tumor Features + UCSF-CAPRA} & 26.8976511 & 6.19E-06 & 7.22E-06 \\
\bottomrule
\end{tabular}
\caption{Results of multivariate log-rank tests comparing the fit of Kaplan-Meier survival curves across models. Reported metrics include test-statistic, p-value, and FDR adjusted p-value.}
\label{tab:s6}
\end{table}

\begin{table}[]
\begin{tabular}{lrlrrr}
\toprule
\multicolumn{6}{c}{\textbf{Supplementary Table 7 (Panel A)}} \\
\midrule
\textbf{Feature} & \multicolumn{1}{l}{\textbf{Adjusted P-value}} & \textbf{Sig} & \multicolumn{1}{l}{\textbf{OR}} & \multicolumn{1}{l}{\textbf{OR\_2.5\_ci}} & \multicolumn{1}{l}{\textbf{OR\_97.5\_ci}} \\
\midrule
shape\_solidity\_median & 5.84E-51 & *** & 1.5936763 & 1.50211585 & 1.69254369 \\
hema\_med\_std & 6.05E-49 & *** & 1.5867459 & 1.49E+00 & 1.68597744 \\
Haralick.IMC1.Mean\_median & 7.07E-46 & *** & 1.59528014 & 1.50E+00 & 1.69964614 \\
size\_area\_std & 1.63E-44 & *** & 1.52582109 & 1.44E+00 & 1.61726074 \\
stdIntensity\_hema\_std & 3.00E-31 & *** & 1.39522233 & 1.32E+00 & 1.47355702 \\
size\_area\_median & 4.22E-31 & *** & 1.42520725 & 1.35E+00 & 1.51069155 \\
stdIntensity\_eosin\_median & 5.70E-30 & *** & 0.68660583 & 6.45E-01 & 0.73081209 \\
size\_major\_axis\_length\_std & 1.97E-25 & *** & 1.35681495 & 1.28427083 & 1.43416294 \\
medianIntensity\_std & 1.24E-24 & *** & 1.36546525 & 1.28981951 & 1.44626022 \\
Haralick.DifferenceEntropy.Range\_median & 7.96E-24 & *** & 0.73653809 & 0.69547586 & 0.7796523 \\
Haralick.IMC2.Range\_std & 3.05E-22 & *** & 1.32596335 & 1.25587441 & 1.40057633 \\
skewIntensity\_hema\_median & 3.23E-22 & *** & 1.41215244 & 1.32137843 & 1.51004992 \\
Haralick.SumAverage.Range\_median & 8.58E-21 & *** & 0.74970952 & 0.70766195 & 0.79398592 \\
fsd3\_median & 1.81E-20 & *** & 0.76053046 & 0.719553 & 0.80340849 \\
medianIntensity\_median & 2.30E-20 & *** & 0.69354163 & 0.6439442 & 0.74645588 \\
shape\_extent\_median & 5.94E-19 & *** & 1.29239471 & 1.22513149 & 1.36385287 \\
fsd6\_std & 4.26E-18 & *** & 1.29232912 & 1.22368056 & 1.36557513 \\
size\_minor\_axis\_length\_std & 4.35E-18 & *** & 1.28733234 & 1.21985163 & 1.35907088 \\
shape\_circularity\_median & 6.67E-18 & *** & 1.28479395 & 1.21757512 & 1.35612147 \\
eosin\_med\_median & 1.14E-17 & *** & 0.72574436 & 0.67691822 & 0.77766425 \\
Haralick.DifferenceVariance.Range\_std & 1.17E-15 & *** & 0.79433864 & 0.75337002 & 0.83737918 \\
Haralick.Correlation.Mean\_std & 1.85E-14 & *** & 1.25319579 & 1.18772761 & 1.3228394 \\
Haralick.IMC1.Mean\_std & 1.87E-14 & *** & 1.25249565 & 1.18718116 & 1.32189554 \\
eosin\_std\_std & 1.00E-13 & *** & 1.25004211 & 1.18379165 & 1.32046932 \\
Haralick.SumAverage.Mean\_std & 3.88E-13 & *** & 1.2441227 & 1.17818124 & 1.31417145 \\
Haralick.DifferenceVariance.Range\_median & 1.03E-12 & *** & 0.78698359 & 0.73924922 & 0.8351205 \\
Haralick.SumAverage.Range\_std & 3.66E-12 & *** & 1.23685568 & 1.17060819 & 1.307167 \\
eosin\_med\_std & 3.79E-12 & *** & 1.23967146 & 1.17260882 & 1.31104136 \\
fsd5\_std & 1.53E-11 & *** & 1.21962179 & 1.15695832 & 1.28594723 \\
ecm\_std\_std & 2.04E-11 & *** & 1.22098495 & 1.1576447 & 1.28819545 \\
minIntensity\_median & 3.70E-11 & *** & 0.77268157 & 0.72046763 & 0.82835807 \\
Haralick.Entropy.Range\_median & 1.08E-10 & *** & 0.82462276 & 0.78174085 & 0.86958321 \\
orientation\_std & 1.35E-10 & *** & 1.2070133 & 1.14574369 & 1.27176859 \\
Haralick.ASM.Range\_std & 1.86E-10 & *** & 1.21405591 & 1.15040835 & 1.2819609 \\
fsd2\_median & 3.93E-10 & *** & 0.72117476 & 0.65536794 & 0.78878563 \\
Haralick.Correlation.Mean\_median & 8.43E-10 & *** & 0.81439136 & 0.76753048 & 0.86384596 \\
Haralick.ASM.Mean\_median & 1.25E-09 & *** & 0.79756015 & 0.74624861 & 0.85095651 \\
Haralick.IMC2.Range\_median & 1.66E-09 & *** & 1.21140613 & 1.14556815 & 1.28137986 \\
maxIntensity\_std & 3.62E-09 & *** & 1.2202555 & 1.15034754 & 1.29483267 \\
Haralick.DifferenceEntropy.Mean\_std & 5.22E-09 & *** & 1.19306251 & 1.13173891 & 1.25803215 \\
Haralick.ASM.Range\_median & 9.54E-09 & *** & 0.82804436 & 0.78169682 & 0.87668369 \\
skewIntensity\_median & 1.26E-08 & *** & 1.2174425 & 1.1465617 & 1.29322257 \\
fsd1\_std & 3.10E-07 & *** & 1.16992035 & 1.11057527 & 1.23260847 \\
\bottomrule
\end{tabular}
\caption{(Panel A) Results from the generalized linear model with patient/slide-level fixed effects for each nuclei feature. P-values were adjusted using Bonferroni correction for multiple comparisons. Significance levels are denoted as *** = adjusted p-value $<0.001$, ** = adjusted p-value $< 0.01$, and * = adjusted p-value $<0.05$. Odds ratios (OR) are reported with 95\% confidence intervals.}
\label{tab:s7}

\begingroup
    \renewcommand{\thetable}{S\arabic{table}A}
    \label{tab:s7A}
\endgroup
\end{table}

\begin{table}[!htbp]\ContinuedFloat
\begin{tabular}{lrlrrr}
\toprule
\multicolumn{6}{c}{\textbf{Supplementary Table 7 (Panel B)}} \\
\midrule
\textbf{Feature} & \multicolumn{1}{l}{\textbf{Adjusted P-value}} & \textbf{Sig} & \multicolumn{1}{l}{\textbf{OR}} & \multicolumn{1}{l}{\textbf{OR\_2.5\_ci}} & \multicolumn{1}{l}{\textbf{OR\_97.5\_ci}} \\
\midrule
gradient\_mag\_mean\_eosin\_std & 6.57E-07 & *** & 1.17861952 & 1.11481631 & 1.24637905 \\
fsd4\_std & 1.78E-06 & *** & 1.16040998 & 1.10172107 & 1.2225332 \\
gradient\_mag\_std\_hema\_std & 4.04E-06 & *** & 1.15852883 & 1.0990499 & 1.22147864 \\
Haralick.SumVariance.Mean\_std & 1.28E-05 & *** & 1.15800212 & 1.09643 & 1.22327444 \\
skewIntensity\_hema\_std & 0.00166249 & ** & 1.1234566 & 1.06511863 & 1.18519974 \\
Haralick.SumEntropy.Range\_median & 0.00551029 & ** & 0.89231585 & 0.84372931 & 0.94349285 \\
Haralick.DifferenceEntropy.Range\_std & 0.00750759 & ** & 0.89944308 & 0.85299275 & 0.94828445 \\
gradient\_mag\_mean\_std & 0.00835043 & ** & 1.11243515 & 1.0544707 & 1.17376512 \\
minIntensity\_eosin\_std & 0.00984315 & ** & 1.10997298 & 1.05275292 & 1.17052374 \\
ecm\_std\_median & 0.01021533 & * & 1.1312003 & 1.06246711 & 1.20464419 \\
fsd1\_median & 0.02820537 & * & 1.10036708 & 1.04452392 & 1.15947344 \\
gradient\_mag\_mean\_eosin\_median & 0.03973418 & * & 0.88077504 & 0.82026767 & 0.94555199 \\
Haralick.Correlation.Range\_std & 0.05152258 &  & 1.09738444 & 1.0407353 & 1.15730126 \\
fsd5\_median & 0.06498476 &  & 1.09631465 & 1.03924972 & 1.15664203 \\
stdIntensity\_median & 0.08592941 &  & 0.90614682 & 0.85441289 & 0.96086533 \\
shape\_solidity\_std & 0.13983769 &  & 0.92010172 & 0.87352473 & 0.96890026 \\
Haralick.IDM.Range\_std & 0.14580725 &  & 0.91963143 & 0.87268458 & 0.96896076 \\
maxIntensity\_median & 0.2083118 &  & 0.90902321 & 0.85461316 & 0.96676314 \\
gradient\_mag\_entropy\_std & 0.21391734 &  & 1.0837304 & 1.028769 & 1.14177778 \\
skewIntensity\_std & 0.23710339 &  & 1.08397388 & 1.02828869 & 1.1428235 \\
gradient\_mag\_mean\_hema\_median & 0.27072877 &  & 1.102872 & 1.03354852 & 1.17704939 \\
minIntensity\_eosin\_median & 0.2718708 &  & 0.89443566 & 0.83046083 & 0.96312211 \\
gradient\_mag\_mean\_median & 1 &  & 1.07451975 & 1.00858347 & 1.14489543 \\
Haralick.IMC1.Range\_std & 1 &  & 1.02298623 & 0.97115662 & 1.07762655 \\
stdIntensity\_eosin\_std & 1 &  & 0.99874708 & 0.94643256 & 1.05395118 \\
skewIntensity\_eosin\_median & 1 &  & 1.02146257 & 0.96531991 & 1.08091859 \\
shape\_circularity\_std & 1 &  & 1.02728108 & 0.97538233 & 1.08198184 \\
fsd3\_std & 1 &  & 1.00885506 & 0.95724904 & 1.06325289 \\
skewIntensity\_eosin\_std & 1 &  & 1.04770111 & 0.99238825 & 1.10617727 \\
Haralick.DifferenceVariance.Mean\_std & 1 &  & 1.01633312 & 0.96106824 & 1.07481321 \\
Haralick.Entropy.Range\_std & 1 &  & 1.00007636 & 0.94894187 & 1.05396639 \\
orientation\_median & 1 &  & 0.93788333 & 0.89075797 & 0.98744637 \\
Haralick.SumEntropy.Range\_std & 1 &  & 1.00428914 & 0.95254238 & 1.05885514 \\
gradient\_mag\_mean\_hema\_std & 1 &  & 1.05637094 & 1.00174185 & 1.11407776 \\
Haralick.SumEntropy.Mean\_median & 1 &  & 0.96274157 & 0.90850589 & 1.02015855 \\
gradient\_mag\_entropy\_median & 1 &  & 1.04746706 & 0.99290198 & 1.10512446 \\
fsd2\_std & 1 &  & 0.99405867 & 0.94381777 & 1.04695117 \\
shape\_extent\_std & 1 &  & 1.0228043 & 0.97119166 & 1.0771943 \\
Haralick.DifferenceVariance.Mean\_median & 1 &  & 0.00013996 & 1.30E-22 & 1.24E+14 \\
fsd6\_median & 1 &  & 1.06811748 & 1.01247864 & 1.12726149 \\
Haralick.ASM.Mean\_std & 1 &  & 1.0138895 & 0.95747501 & 1.07420277 \\
fsd4\_median & 1 &  & 1.00960457 & 0.95809841 & 1.06389238 \\
Haralick.Correlation.Range\_median & 1 &  & 0.98099981 & 0.92482779 & 1.04055504 \\
\bottomrule
\end{tabular}
\caption{(Panel B) Results continued from the generalized linear model with patient/slide-level fixed effects for each nuclei feature. P-values were adjusted using Bonferroni correction for multiple comparisons. Significance levels are denoted as *** = adjusted p-value $<0.001$, ** = adjusted p-value $< 0.01$, and * = adjusted p-value $<0.05$. Odds ratios (OR) are reported with 95\% confidence intervals.}
\begingroup
    \renewcommand{\thetable}{S\arabic{table}B}
    \label{tab:s7B}
\endgroup
\end{table}

\begin{table}[]
\centering
\begin{tabular}{lrlrrr}
\toprule
\multicolumn{6}{c}{\textbf{Supplementary Table 8}} \\
\midrule
\textbf{Feature} & \multicolumn{1}{l}{\textbf{Adjusted P-value}} & \textbf{Sig} & \multicolumn{1}{l}{\textbf{OR}} & \multicolumn{1}{l}{\textbf{OR\_2.5\_ci}} & \multicolumn{1}{l}{\textbf{OR\_97.5\_ci}} \\
\midrule
alignment\_max & 9.30E-13 & *** & 0.80585946 & 0.76176328 & 0.85164257 \\
intensity\_summ & 1.03E-12 & *** & 1.28549296 & 1.20480345 & 1.37200024 \\
coh & 2.08E-12 & *** & 1.2838765 & 1.20275395 & 1.37095112 \\
coh\_w & 2.08E-12 & *** & 1.2838765 & 1.20275395 & 1.37095112 \\
collagen\_volume & 4.24E-12 & *** & 1.27410609 & 1.19497751 & 1.35884846 \\
intensity\_median & 7.59E-10 & *** & 1.24535201 & 1.16771778 & 1.32855593 \\
length\_max & 1.38E-09 & *** & 0.81988595 & 0.77195318 & 0.86894447 \\
strightness\_max & 6.06E-09 & *** & 0.84206761 & 0.79849528 & 0.88783125 \\
intensity\_max & 1.49E-07 & *** & 1.20263987 & 1.13056496 & 1.27958449 \\
length\_var & 7.90E-06 & *** & 1.16365564 & 1.09848011 & 1.23307096 \\
thickness\_min & 3.38E-05 & *** & 1.15981755 & 1.09262525 & 1.23140492 \\
length\_min & 8.12E-05 & *** & 1.14490981 & 1.08202056 & 1.21177197 \\
fractal\_dim & 0.00018378 & *** & 1.15440652 & 1.08477889 & 1.22884441 \\
alignment\_min & 0.00175196 & ** & 1.13932162 & 1.07223587 & 1.2186017 \\
length\_mean & 0.00268078 & ** & 1.12163495 & 1.05900265 & 1.18820281 \\
thickness\_max & 0.00332224 & ** & 0.89775788 & 0.84962021 & 0.94808438 \\
thickness\_var & 0.00625118 & ** & 1.12043792 & 1.05501577 & 1.19012229 \\
angle\_mean & 0.00764005 & ** & 1.10143349 & 1.04577455 & 1.16016421 \\
strightness\_mean & 0.03537363 & * & 0.9173012 & 0.87048002 & 0.96650454 \\
thickness\_mean & 0.04197028 & * & 1.10813523 & 1.04036502 & 1.18053523 \\
strightness\_min & 0.11082502 &  & 1.09495421 & 1.03171789 & 1.16584529 \\
strightness\_var & 0.27438263 &  & 1.07174837 & 1.01724793 & 1.12954511 \\
angle & 0.41259102 &  & 0.93718649 & 0.88975268 & 0.98702976 \\
orientation & 0.45437906 &  & 1.06603103 & 1.0122018 & 1.12285398 \\
alignment\_mean & 0.57255499 &  & 0.93999348 & 0.89212004 & 0.99003421 \\
angle\_min & 0.76296785 &  & 1.06135053 & 1.00721105 & 1.11889111 \\
angle\_var & 1 &  & 0.94649327 & 0.89859715 & 0.99687971 \\
alignment\_var & 1 &  & 1.02279899 & 0.97119707 & 1.07732973 \\
angle\_max & 1 &  & 0.94537751 & 0.89714212 & 0.99612442 \\
\bottomrule
\end{tabular}
\caption{Results from the generalized linear model with patient/slide-level fixed effects for each collagen feature. P-values were adjusted using Bonferroni correction for multiple comparisons. Significance levels are denoted as *** = adjusted p-value $<0.001$, ** = adjusted p-value $< 0.01$, and * = adjusted p-value $<0.05$. Odds ratios (OR) are reported with 95\% confidence intervals.}
\label{tab:s8}
\end{table}

\begin{table}[]
\centering
\begin{tabular}{llrrrr}
\toprule
\multicolumn{6}{c}{\textbf{Supplementary Table 9}} \\
\midrule
\textbf{Feature Set} & \textbf{Model} & \multicolumn{1}{c}{\textbf{AUROC}} & \multicolumn{1}{c}{\textbf{Accuracy}} & \multicolumn{1}{c}{\textbf{Sensitivity}} & \multicolumn{1}{c}{\textbf{Specificity}} \\
\midrule
\textit{Nuclei} & \textit{Random Forest} & 0.686 & 0.637 & \textbf{0.584} & 0.692 \\
\textit{Nuclei} & \textit{XGBoost} & 0.681 & 0.64 & 0.522 & 0.761 \\
\textit{Nuclei} & \textit{naive Bayes} & 0.652 & 0.609 & 0.607 & 0.612 \\
\textit{Nuclei} & \textit{Neural Network} & 0.725 & \textbf{0.672} & \textbf{0.584} & 0.763 \\
\textit{Collagen} & \textit{Random Forest} & 0.519 & 0.511 & 0.272 & 0.759 \\
\textit{Collagen} & \textit{XGBoost} & 0.522 & 0.523 & 0.233 & 0.823 \\
\textit{Collagen} & \textit{naive Bayes} & 0.473 & 0.468 & 0.551 & 0.382 \\                                                                              
\textit{Collagen} & \textit{Neural Network} & 0.505 & 0.473 & 0.526 & 0.419 \\
\textit{Combined} & \textit{Random Forest} & 0.697 & 0.642 & 0.609 & 0.676 \\
\textit{Combined} & \textit{XGBoost} & 0.696 & 0.648 & 0.563 & 0.735 \\
\textit{Combined} & \textit{naive Bayes} & 0.598 & 0.577 & 0.385 & 0.774 \\
\textit{Combined} & \textit{Neural Network} & \textbf{0.728} & 0.67 & 0.527 & \textbf{0.817} \\
\bottomrule
\end{tabular}
\caption{Performance of machine learning models trained on hand-crafted feature sets (nuclei, collagen, and combined) evaluated on the hold-out test set. The highest value for each performance metric is shown in bold.}
\label{tab:s9}
\end{table}

\begin{sidewaystable}[p]
\centering
\scriptsize
\setlength{\tabcolsep}{4pt}
\renewcommand{\arraystretch}{1.1}

\begin{tabularx}{\textheight}{>{\raggedright\arraybackslash}p{5.2cm} *{6}{>{\centering\arraybackslash}X}}
\toprule
\multicolumn{7}{c}{\textbf{Supplementary Table 10}}\\
\midrule
\textbf{Pathologist Survey Results} & \textbf{Recall} & \textbf{Precision} & \textbf{Accuracy} & \textbf{Cytoplasm Percentage} & \textbf{Nuclei Percentage} & \textbf{ECM Percentage} \\
\midrule
PROTAS & 0.88 & 0.64 & 0.70 & -- & -- & -- \\
Pathologist 1 & 0.59 & 0.55 & 0.56 & 37\% & 49\% & 28\% \\
Pathologist 2 & 0.49 & 0.60 & 0.59 & 50\% & 84\% & 28\% \\
Pathologist 3\textsuperscript{*} & 0.61 & 0.51 & 0.52 & 79\% & 7\%  & 83\% \\
Pathologist 4 & 0.29 & 0.64 & 0.57 & 91\% & 100\% & 87\% \\
Majority Vote (All Pathologists) & 0.51 & 0.54 & 0.55 & -- & -- & -- \\
Majority Vote (Only Prostate Path) & 0.49 & 0.65 & 0.62 & -- & -- & -- \\
\bottomrule
\end{tabularx}

\vspace{0.3em}
\begin{minipage}{\textheight}
\scriptsize
\textit{Note:} \textsuperscript{*} Non-prostate pathologist.
\end{minipage}

\caption{Performance metrics (recall, precision, and accuracy) for PROTAS, each pathologist, and majority-vote aggregations. Percentages indicates the proportion of patches where cytoplasm, nuclei, and extracellular matrix (ECM) were marked as informative for pathologist decision making.}
\label{tab:s10}
\end{sidewaystable}

\begin{table}[]
\centering
\begin{tabular}{lc}
\toprule
\multicolumn{2}{c}{\textbf{Supplementary Table 11}} \\
\midrule
\textbf{Cohen's Kappa} & \textbf{kappa} \\
\textit{Model vs Pathologist 1} & 0.143 \\
\textit{Model vs Pathologist 2} & 0.045 \\
\textit{Model vs Pathologist 3} & -0.023 \\
\textit{Model vs Pathologist 4} & 0.042 \\
\textit{Pathologist 1 vs Pathologist 2} & 0.269 \\
\textit{Pathologist 1 vs Pathologist 3} & 0.03 \\
\textit{Pathologist 1 vs Pathologist 4} & 0.013 \\
\textit{Pathologist 2 vs Pathologist 3} & 0.208 \\
\textit{Pathologist 2 vs Pathologist 4} & 0.279 \\
\textit{Pathologist 3 vs Pathologist 4} & 0.255 \\
\bottomrule
\end{tabular}
\caption{Cohen's $\kappa$ agreement scores between the model and each pathologist, and between pathologist pairs.}
\label{tab:s11}
\end{table}
%\end{comment}

\end{document}